 \documentclass[final,3p,times]{elsarticle}

\biboptions{sort&compress}

\usepackage{epsfig,rotating,graphics}
\usepackage{amsmath}
\usepackage{amsfonts}
\usepackage{amssymb}
\usepackage{comment}

\usepackage{amsthm}
\usepackage{fontenc}
\usepackage{graphicx}
\usepackage{graphics}
\usepackage{blindtext}
\usepackage{lipsum}
\usepackage{tikz}
\usepackage{soul}
\usepackage[normalem]{ulem}
\def\be{\begin{eqnarray} &&}

\def\ee{\end{eqnarray}}

\newcommand{\bfi}{\begin{figure}}
\newcommand{\efi}{\end{figure}}
\newcommand{\delete}[1]{{\blue{\sout{#1}}}}
\newcommand{\replace}[2]{\red{\sout{#1}}\blue{#2}}
\usepackage{multirow}
\usepackage{hhline}
\usepackage{tikz}
\usepackage{makecell}
\usepackage{comment}

\usepackage{caption}
\newcommand{\mc}[3]{\multicolumn{#1}{#2}{#3}}
\definecolor{tcA}{rgb}{0,0,0}
\usepackage{array}
\newcolumntype{P}[1]{>{\centering\arraybackslash}p{#1}}

\newcommand{\red}[1]{\textcolor[rgb]{1,0,0}{#1}}

\newcommand{\green}[1]{{\textcolor[rgb]{0,0.7,0.}{#1}}}
\newcommand{\lightbrown}[1]{{\textcolor[rgb]{0.71, 0.4, 0.11}{#1}}}

\newcommand{\blue}[1]{{\textcolor[rgb]{0,0,1}{#1}}}
\newcommand{\pink}[1]{{\textcolor[rgb]{1,0,0.5}{#1}}}

\newcommand{\gianni}[1]{{ \green{{\bf Gianni}:\ {#1}}}}

\newcommand{\fede}[1]{{ \red{{\bf Fede}:\ {#1}}}}

\newcommand{\matteo}[1]{{ \pink{{\bf Matteo}:\ {#1}}}}
\newcommand{\emanuele}[1]{{ \lightbrown{{\bf Emanuele}:\ {#1}}}}

\journal{Physics Letters B}

\begin{document}

\begin{frontmatter}



\title{Double parton scattering off nuclei and the double-EMC effect}


\author[first,sixth]{Filippo Fornetti}
\affiliation[first]{organization={Dipartimento di Fisica e Geologia, Universit\`a  degli Studi di Perugia and INFN Sezione di Perugia, Via A. Pascoli, Perugia, Italy}
        }
\affiliation[sixth]{organization={Institute of Modern Physics, Chinese Academy of Sciences, Lanzhou 73000, China}
        }

        \author[second]{Federico Alberto Ceccopieri}
\affiliation[second]{organization={Université de Liège, Interactions Fondamentales en Physique et en Astrophysique, Liège, Belgium}
        }

        \author[third]{Emanuele Pace}
\affiliation[third]{organization={Universit\`a di Roma Tor Vergata and INFN Sezione di Roma Tor Vergata,
Via della Ricerca Scientifica 1, 00133 Rome, Italy}
        }

       \author[fourth]{Matteo Rinaldi \corref{cor1}}
\affiliation[fourth]{organization={Istituto Nazionale di Fisica Nucleare
Sezione di Perugia, Via A. Pascoli, Perugia, Italy}
        }

        \cortext[cor1]{Corresponding author}

\ead{matteo.rinaldi@pg.infn.it}

       \author[fifth]{Giovanni Salm\`e}
\affiliation[fifth]{organization={Istituto  Nazionale di Fisica Nucleare, Sezione di Roma, Piazzale A. Moro 2,
00185 Rome, Italy}
        }

\begin{abstract}

   The double-parton scattering process  off light and heavy nuclei is proposed as an {effective} tool to open a novel window for accurately studying the effects of nuclear binding  on the internal structure of the nucleon, in the spirit of the European Muon Collaboration (EMC) effect. The key ingredient is the {actual calculation of} {\em nuclear double-parton distributions} that contain information about   double-parton distributions and generalized-parton distributions {of the nucleon}, folded with nuclear light-front distributions.  {It allows to introduce} a new quantity,  {{which we call}} {\em double-EMC ratio} {{that}} appears to be much more sensitive  than the {standard} EMC effect to  {both} the nucleon partonic structure and    to the nuclear {binding}. Hence, it could help solving
long-standing questions about how nucleons change when {they are} bound in atomic nuclei.

\end{abstract}




\end{frontmatter}




\section{Introduction}
   In recent  decades,
   significant efforts have been devoted to characterize   signatures of multi-parton interactions (MPI) in hadronic collisions. Notably, it has been found that {MPI} play a  crucial role in  studies of Standard Model processes and beyond, when high accuracy is required (see, e.g. \cite{Paver:1982yp,4a,book}). In particular,
a special emphasis has been {paid to}   
double-parton (hard) scattering (DPS), which is the simplest process within the set of  {multi-parton scatterings}.  It can be treated by applying the perturbation theory, if 
hard scales are involved in both parton-parton scatterings.
Although  DPS is generally suppressed w.r.t.  single-parton scattering (SPS),  it is  possible to properly choose specific final states for which DPS is {comparable} with the SPS {background}. 
{In this regard}, same-sign ($ss$) $WW$ production is  a gold-plated channel, due to charge conservation leading to an enhancement of the DPS signal w.r.t. the SPS signal \cite{Gaunt:2010pi}. Noteworthy, the process has  been recently  measured at the LHC \cite{CMS:2022pio,ATLAS:2025bcb}.
Moreover, DPS is relevant in double $J/\psi$ hadron production
in some regions of the phase space (see e.g. \cite{LHCb:2016wuo,Borschensky:2016nkv,Lansberg:2014swa}), and it  seems to play a  
central role in elucidating the 
{interpretation} of the first observation of triple $J/\psi$ production \cite{CMS:2021qsn}.

From a theoretical point of view, 
 the investigation of DPS offers the possibility to access double-parton distributions  (DPDs)  \cite{gauntevo,4a} {(see  also Ref. \cite{Diehl:2018wfy} for further studies on general properties of  DPDs)}, which encode  the probability of finding {two  partons} {in a nucleon}  with given longitudinal momentum fractions, $x_1$ and $x_2$,
 at relative transverse distance 
 ${\bf y}_\perp$. 
 Assuming the Fock-expansion of the hadron state,
 one can introduce the DPDs  and  study their essential role  in {generating},  e.g., a multidimensional picture of the proton \cite{4a,Rinaldi:2018slz,jhepc,Bali:2018nde,man,noi1}, basically different from that obtained via single-parton generalized parton distribution functions  {(GPDs)} \cite{Diehl:2003ny} or 
 transverse-momentum dependent parton distribution functions {(TMD)} \cite{Angeles-Martinez:2015sea}. 
 In order to enhance the DPS rate, nuclear targets  represent the optimal choice (see, e.g.,  Refs. \cite{Treleani:2012zi,Strikman:2001gz,Cattaruzza:2005nu,Blok:2012jr,Ceccopieri:2025edn} for theoretical studies   and Refs. \cite{Abreu:2023wwg,dEnterria:2017yhd,dEnterria:2012jam,dEnterria:2013mrp,dEnterria:2014lwk,dEnterria:2016yhy,dEnterria:2016ids,Blok:2020oce,Blok:2019fgg,CMS:2024wgu} for phenomenological analyses).  For heavy nuclei, the number of events is increased by almost a factor $A^{4/3}$ with $A$ the number of nucleons in the nucleus \cite{Strikman:2001gz,Blok:2012jr,Blok:2019fgg},
{ {therefore} the planned $pA$-run foreseen at the LHC \cite{dEnterria:2025jgm}  has the potential to further enhance the DPS program in the nuclear sector.}

Our aim is to explore,  whether and to what extent nuclear DPS could provide novel insights into the partonic structure of nuclei in the valence region {i.e. $0.3\le x\le 0.7$}. 
To this end, we calculate  nuclear DPDs, retaining only the valence component of the nuclear Fock state \cite{Ceccopieri:2025edn}, {and disregarding both target-mass corrections {(TMC)} and higher-twists in the nucleon DPDs (the investigation of such corrections is deferred to future works, since the current  goal is  to illustrate the backbone of our proposal)}. {In this work}, we evaluate {nuclear DPDs} for $^2$H{, $^3$He} and $^4$He, as well as  $^3$H {(see supplemental material (SM))  by using  refined calculations of the nuclear wave functions,  based on realistic interactions  (see Refs. \cite{Kievsky:1994mxj,Kievsky:1995uk}) {and of the light-cone momentum distributions (LCMDs)} given in a Poincar\'e-covariant framework (see e.g. \cite{Pace:2022qoj,Fornetti:2023gvf}). For providing  a general overview  of a wide  $A$-range, {we calculate} heavy-nuclei DPDs  by adopting the light-front holographic model (LFHM) of Ref. \cite{Kim:2022lng}.
Then, we {introduce and} investigate novel ratios  of nuclear {DPDs}, which we call \textit{double-EMC ratios}, to be interpreted as the two-body counterpart {in DPS processes} of the conventional EMC ratio. {Finally, we show that while distinct models of nucleon structure, adopted in  both DIS and DPS (see Refs. \cite{faccioli,noiprl,Kim:2022lng}), {can} provide substantially equivalent {{descriptions}} for the single EMC ratios, they {{yield remarkably}}   different  predictions for the double-EMC ratios.}
Therefore these quantities could   { lead to gain} fresh information on the partonic structure of bound nucleons, shedding further light on the dynamics underlying the EMC effect.

\section{Nuclear double-parton distributions}
 Nuclear DPDs are defined following
 Refs. \cite{Blok:2012jr,Ceccopieri:2025edn}, {where {the partonic structure of  single  nucleons is folded with the structure of nuclear body, calculated by retaining} only nucleonic degrees of freedom (dofs). 
 As {extensively} discussed in, e.g., Refs. \cite{Blok:2012jr,Blok:2019fgg,Strikman:2001gz,dEnterria:2017yhd,Ceccopieri:2025edn}, for $pA$ and 
     $\gamma^* A$ collisions, the cross-section {for DPS} has two contributions: $i)$ DPS1,  generated by two partons both belonging to  the {\em same }nucleon in the nucleus; $ii)$ DPS2, generated by two partons  belonging to  {\em different} nucleons.  {In particular for heavy nuclei,} the cross-section  scales with {the number of nucleons}  as $A ~\sigma_{DPS}^{pp}$ for the nuclear  DPS1, and  as $\sigma_{DPS2}^{pA} \propto A^{1/3}~ \sigma_{DPS1}^{pA}$   for the nuclear DPS2 \cite{Blok:2019fgg}
  (see, e.g., Refs. \cite{Fedkevych:2020cmd,Huayra:2019iun}, for  recent theoretical analyses  in  $pp$ and $pA$ collisions).

In this letter, we focus on the DPS1 contribution, that brings {peculiar} information on the single nucleon structure \cite{Blok:2012jr,Ceccopieri:2025edn},  
 deferring to future analyses the detailed investigation of the  DPS2 contribution that opens a window on GPDs in nuclei. Further information on the DPS2 mechanism can be found in, e.g., Ref. \cite{Ceccopieri:2025edn}, where it is pointed out that DPS2, being dependent on {the product of} nucleon GPDs,
populates the $\{x_1,x_2\}$ {plane} differently from the DPS1 contribution.
In order to estimate 
{the relative weight of the DPS1 and DPS2 mechanisms}, we follow the line of Ref. \cite{Ceccopieri:2025edn}}, where the Gaunt-Stirling (GS)
 sum rules for DPDs \cite{gauntevo}
 have been applied {separately to the DPS1 and the DPS2 contributions.} {The comparison of  the DPS1 contribution to  the full sum rules,  
   confirms that {in general} the DPS1 mechanism is not the dominant one, but, depending on the chosen flavors, it can compete with the DPS2 one, in particular for light-nuclei (see  Sect. S1 in SM). 
Remarkably, {the DPS2 contribution could be in principle subtracted from the total cross-seciton, since workable estimates of the DPS2  term   can be obtained from the existing  phenomenological parametrizations of GPDs  and the knowledge of the two-body nuclear density. The latter can be evaluated when the nuclear wave function is known or, particularly for heavy nuclei,   a factorized  form based on the one-body nuclear density is {{adopted}} \cite{Strikman:2001gz,Cattaruzza:2005nu,Blok:2012jr,Blok:2020oce,Blok:2019fgg,Blok:2020ckm,Salvini:2013xpz,Cattaruzza:2004qb}. }


 In impulse approximation (IA), the DPS1 contribution to the nuclear DPD is defined in the nucleus rest frame, where the nucleus has light-front (LF) four-momentum $P^\mu_A\equiv \{P^\pm_A=M_A {/\sqrt{2}}, {\bf P}_{A;\perp}=0\}$  with $M_A$ the nucleus mass, as follows  {(see also Ref. \cite{Ceccopieri:2025edn})}:
\begin{align}
    \label{Eq:DPD}
  D_{q_{1} q_{2}}^{A,1}\left(x_{1}, x_{2}, \bf{k}_{\perp}\right)&=\sum_{\tau=p, n} \int_{\xi_{min}}^{1} d \xi ~\frac{{\bar{\xi}^2}}{\xi^{2}}~ \bar \rho^A_{\tau}(\xi)
    D_{q_1, q_2}^{\tau}\left(\frac{x_{1}}{\xi} \bar{\xi}, \frac{x_{2}}{\xi} \bar{\xi}, \bf{k}_\perp\right) ~,
\end{align}
where  {$x_i=p^+_i/P^+_A \times (1/\bar \xi)$  are the parton longitudinal-momentum ratios w.r.t.  the nucleus momentum, {such that}  $\sum_i x_i=1/\bar \xi=M_A/m$, with   $p_i^\mu\equiv\{p^+_i= x_i \bar \xi P_A^+ , p^-_i,{\bf p}_{i;\perp}\}$ the $i$-th parton four-momentum and $m$ the nucleon mass} {{(the LF components  are defined as $a^\pm = (a^0 \pm a^3)/\sqrt{2}$)}}. {The function $D_{q_1, q_2}^{\tau}(z_1,z_2,k_\perp)$ is the nucleon DPD, with $z_i = p_i^+/P_N^+=x_i~\bar \xi/\xi $ the parton longitudinal-momentum ratios w.r.t.  the parent nucleon momentum, which is given by  $P_N^\mu\equiv\{ {\xi P^+_A},P^-_N,{\bf P}_{N;\perp}\}$},  and ${\bf  k}_\perp$  the conjugate variable to {the relative distance} ${\bf y}_\perp$.   
 The function  $D_{q_1, q_2}^{\tau}(z_1,z_2,{\bf k}_\perp=0) $   gives  the probability density for finding partons with momentum fractions  (flavors) $z_1$ ($q_1$) and $z_2$ ($q_2$), respectively. Finally, $\xi$ is the longitudinal-momentum fraction carried by a
nucleon in the nucleus, {$\xi_{min} = (x_1+x_2)\bar \xi$} and $\bar \rho_\tau^A(\xi)$  the
one-body {{LCMD}} of the $\tau$
nucleon in the nucleus, {normalized to $A$} {(see, e.g., Refs. \cite{Pace:2022qoj,Ceccopieri:2025edn} and references  therein)}. For the sake of a light notation, the dependence upon the hard scales has been omitted {here and in the following equations.}

 In our calculations, where only nucleonic dofs have been taken into account, {the LCMD}  satisfies automatically the kinematic support  and  both baryon-number and longitudinal-momentum sum rules, thanks to the Poincar\'e covariance of our light-front framework \cite{Pace:2020ned,Alessandro:2021cbg,Pace:2022qoj,Fornetti:2023gvf,Pace:2001vg}.
One should notice that Eq. (\ref{Eq:DPD}) is the generalization of the {{expression for the nuclear PDF}} obtained in IA, {which reads}:
\begin{align}
\label{Eq:npdf}
    d^A_q(x) = \sum_{\tau=p, n} \int_{ {\xi_{m}}}^{1} \hspace{-0.3 cm} d \xi ~\frac{\bar \xi}{\xi}~ \bar \rho^A_{\tau}(\xi) ~d^\tau_q \left(\frac{x}{\xi}\bar \xi \right)~,
\end{align}
where $d^\tau_q(x)$ is the  PDF of the parton $q$ in the $\tau$ nucleon {and $\xi_{m}=x \bar \xi$}. {Given the exploratory nature of our study, we chose to keep the approach as simple as possible,   i.e.  for now {{TMC}} and higher twists {are disregarded}, since their treatment  at the level of nucleon DPDs deserves dedicate efforts.}
{Important to note that} the
LCMDs  {for light-nuclei} have been obtained from nuclear wave-functions \cite{Kievsky:1994mxj,Kievsky:1995uk} corresponding to   two-nucleon
Av18 \cite{Wiringa:1994wb}  and  three-nucleon Urbana IX \cite{Pudliner:1995wk}  realistic potentials. 
So far the possibility of revealing new insights into the origin of the EMC effect from  nuclear DPS is a conjecture. In order to quantitatively explore the potential of our approach, it is useful to introduce the EMC ratio \cite{EuropeanMuon:1983wih} in terms of DIS cross-sections:
 \begin{align}
    R^A_{EMC}(x) &= \dfrac{d \sigma^A_{DIS}/
   \big\{Z d \sigma^p_{DIS}(x) +(A-Z) d \sigma^n_{DIS}(x)\big\}}{d \sigma^{^2H}_{DIS}/
   \big\{ d \sigma^p_{DIS}(x)+d \sigma^n_{DIS}(x)\big\}}
  =\dfrac{  F_2^{A}(x)/
   \big\{Z F_2^{p}(x)+(A-Z)F_2^{n}(x) \big\}
    }{ F_2^{^2H}(x) /
   \big\{ F_2^{p}(x)+F_2^{n}(x) \big\} } ~,
      \label{Eq:EMC}
   \end{align}
where $d \sigma^{A(^2H)}_{DIS}$ is the nucleus (deuteron) DIS cross-section, $d \sigma^{p(n)}_{DIS}(x)$ the one of the proton (neutron) and  the {final ratio}   is expressed in terms of nucleon, $F_2^{p(n)}(x)$, and nuclear, { $ F_2^{A} (x)$,} structure functions (SFs):
 \begin{align}
 \label{Eq:SF}
     F_2^{A(^2H)} (x) = \sum_{q_i}  e_{q_i}^2 x~ d_{q_i}^{A(^2H)}(x).
 \end{align}

{In the present {{letter}}, we propose to study, for the first time, ratio of DPS cross-sections in order to address to what extent DPS could be used to shed some light on {{the effects of binding on the partonic
structure of nucleons in nuclei}}.  To this aim, {as an exploratory example,} let us {consider} the DPS cross-section}
for the process $pA\rightarrow W^\pm W^\pm$.  {This reaction} is the gold channel for investigating DPSs,
 {since} the SPS background can produce 
a doubly-charged final state only via higher order processes in the strong coupling and can be subtracted by vetoing the additional jets accompanying the $ssWW$. 
For the seek of concreteness,  we investigate the high-energy collision of a proton of 4-momentum $p_{in}$ and a nucleus with atomic number $A$ to produce a $W^+W^+$ pair in the final state (analogous results  for the $W^- W^-$ case can be easily obtained) \cite{noiprl,Kulesza,Blok:2019fgg,dEnterria:2012jam}: 
\begin{equation}
A(P_A) + p(p_{in}) \rightarrow W^+(q_1) + W^+(q_2) +X
\end{equation}
We assume the nucleus   moving along {$+\hat{\bf e}_z$}  and the proton moving along {$-\hat{\bf e}_z$}, in the center of mass (CM) frame of the {incoming} proton  and of the {bound} nucleon $N$   {with 4-momentum $P_N$}. Then, the {square,} per-nucleon CM energy of the collision is 
given by $(p_{in}+P_N)^2=s_{pN}$.
In  DPS, the two $W$'s are produced   by the annihilation 
of a pair of partons from the 
 nucleus, with momenta $\{p_1,~ p_2\}$ and longitudinal momentum fractions $\{x_1,~ x_2\}$, and a pair from the proton, with momenta $\{p_3, ~p_4\}$ and longitudinal momentum fractions $\{z_1, ~z_2\}$. Namely:
\begin{eqnarray}
    i(p_1)+j(p_3) \rightarrow W(q_1); ~~~
    k(p_2)+l(p_4) \rightarrow W(q_2). 
    \label{WW}
\end{eqnarray}

Here $i,j,k,l$ indicate flavor combination relevant for the process:
$u\bar{d}$ annihilation is {involved in } $W^+$ production and  $d\bar{u}$ for $W^-$. Notice that  Cabibbo-suppressed channels have been omitted for clarity. 
{From } the above kinematics, both annihilations must give  objects whose invariant {masses} are  
  $  (p_1+p_3)^2=q_1^2=M_W^2$ and  
  $  (p_2+p_4)^2=q_2^2=M_W^2$, 
{that eventually lead  to}  {$x_1 z_1 =M_W^2/s_{pN}$} and {$x_2 z_2 =M_W^2/s_{pN}$}. These products are fixed once the boson mass and the CM energy of the collisions are set. 
{  We may identify some relevant  regions in the $ssWW$ phase space according to  {{the two boson}} rapidities, 
$ Y_1=1/2 \ln x_1/z_1$ and $ Y_2=1/2 \ln x_2/z_2$, respectively (see discussion in S2 of the SM):}
i) if $x_1 \ll z_1$ and $x_2 \ll z_2$,  the $ssWW$  are produced along the proton  moving-direction ($-\hat{\bf e}_z$) with large and negative rapidities; ii)
if $x_1 \gg z_1$ and $x_2 \gg z_2$, both charged $Ws$ are produced along the nucleus  moving-direction ($+\hat {\bf e}_z$) with large and positive rapidities; 
iii) if $x_1 \ll z_1$ and $x_2 \gg z_2$ or $x_1 \gg z_1$ and $x_2 \ll z_2$, the two  $ssWW$ are produced in opposite hemispheres.
The second configuration is particularly important since it allows the selection of $u$-valence pair in the nucleus, {being  $x_1 \gg z_1$ and $x_2 \gg z_2$}. 
Therefore in such a phase space region the $W^+W^+$ cross section can be schematically written as {\cite{Kulesza,noiprl,dEnterria:2012jam,Diehl:2015bca}  } 
\begin{align}
d \sigma^{A}_{DPS} = {1 \over 2} 
  S^{pA}_{\bar d \bar d u_v u_v}\!\!(z_1,z_2,x_1,x_2)
~ d \hat \sigma_{\bar d u_v}^{W^+} \, d \hat 
\sigma_{\bar d u_v}^{W^+}\,,
\label{uno}
\end{align}
where
\begin{align} 
\label{Eq:SpA}
    S^{pA}_{\bar d \bar d u_v u_v}\!\!(z_1,z_2,x_1,x_2) =
\int {d^2 {\bf k}_{\perp}\over (2 \pi)^2} 
 D^p_{\bar d \bar d}\left(z_1, z_2,- {\bf k}_{\perp} \right) D^A_{u_v u_v}\left(x_1, x_2, {\bf k}_{\perp}\right),
\end{align}
is the soft part of the cross-section, and  
$d \sigma$ and $d \hat{\sigma}$ indicate the hadronic  and partonic cross-sections, respectively, kept differential in the relevant
variables. 
Since the function  in Eq. \eqref{Eq:SpA} depends on the nuclear DPD, one  can tentatively build a ratio,  analogous to Eq. \eqref{Eq:EMC}, but using the {DPS} cross-section in Eq. \eqref{uno}.  We call  {this quantity} \textit{double-EMC ratio} {given} for the  present specific flavor combination by:

\begin{align}
   \label{Eq:2EMC}
R^{2,A}(z_1,z_2,x_1,x_2) &\equiv \dfrac{d \sigma_{DPS}^A/
   \big\{Z d \sigma_{DPS}^p \big\}}{d \sigma_{DPS}^{^2 H}/
   \big\{ d \sigma_{DPS}^p \big\}}
= \dfrac{ S^{pA}_{\bar d \bar d u_v u_v}\!\!(z_1,z_2,x_1,x_2)/
   Z  }{ S^{p ^2H}_{\bar d \bar d u_v u_v}\!\!(z_1,z_2,x_1,x_2)
   }. 
\end{align}

It is very gratifying that, adopting standard assumptions, it is possible to obtain a ratio between cross-sections that is  sensitive to DPDs. {In fact}, the function $S^{pA}_{\bar d \bar d u_v u_v} $ is a folding of  proton and nuclear  DPDs (differently from the DIS case, where  the nuclear SF enters). 
In this first investigation, we will retain  only the DPS1 contribution to $S^{pA}_{\bar d \bar d u_v u_v}$, {i.e. $S^{pA,1}_{\bar d \bar d u_v u_v}$, }leaving the
calculation of DPS2, which needs the nuclear double {LCMDs} (see e.g., Ref. \cite{Ceccopieri:2025edn} where the deuteron case has been already investigated), to future analyses. {  However, as already mentioned, one can achieve  a realistic quantitative estimate of the DPS2 contribution by only using standard assumptions {and approximations} \cite{Strikman:2001gz,Cattaruzza:2005nu,Blok:2012jr,Blok:2020oce,Blok:2019fgg,Blok:2020ckm,Salvini:2013xpz,Cattaruzza:2004qb}. Within this choice,
inserting Eq. \eqref{Eq:DPD} in Eq. \eqref{Eq:SpA}, one gets the soft part entering  Eq. \eqref{Eq:2EMC}:
   {\begin{align}
\label{Eq:2EMCa}
    R^{2,A,1}(z_1,z_2,x_1,x_2) 
    = \dfrac{1}{Z} \dfrac{ S^{pA,1}_{\bar d \bar d u_v u_v}\!\!(z_1,z_2,x_1,x_2)}{ S^{p ^2H,1}_{\bar d \bar d u_v u_v}\!\!(z_1,z_2,x_1,x_2)}.   
\end{align}}
{{Then in}} IA, for  the double-EMC ratio one recovers {{a form analogous}}  to the standard EMC  ratio,  expressed in terms of convolutions of  nucleon soft functions and the nuclear LCMD. 

\section{Models of nucleon DPDs}
 Through the double-EMC ratio,  Eq. \eqref{Eq:2EMCa}, our analysis aims  {to gain information}  on general features related to the  genuinely partonic dynamics underlying  the nuclear EMC effect in the valence region  $0.3 \leq x \leq 0.7$. {To this end, in } Eq. \eqref{Eq:2EMCa}, we adopted   two  {sharply different} classes of free-nucleon DPD models  in the valence region: i) a
\textit{factorized} {one}, where the DPDs are built from the product of   PDFs  times  a function of $k_\perp$ \cite{gauntevo,Golec-Biernat:2022wkx,Golec-Biernat:2014bva,Kim:2022lng}, {neglecting} unknown non-perturbative  double-parton correlations (DPCs)   (see Sect. S3 in SM for details), and ii) {a} \textit{fully correlated} {relativistic} constituent quark model (CQM), where two-body  correlations are encoded in DPDs \cite{noij1,jhepc},  which plainly differs from the product of corresponding PDFs times a universal function of $k_\perp$  (see Sect. S4 in SM).

 The {\em factorized} DPD {has a schematic form:},  
 $D^p_{ij}(x_1,x_2,k_\perp) \sim  d^p_i(x_1)d^p_j(x_2) g(k_\perp)$, where $g(k_\perp)$ is a function that {simplifies} in DPDs ratios.
{Some physical constraints have been taken into account, as i) 
the correct support, i.e. $D^p_{ij}(x_1,x_2,\textbf{k}_\perp)=0$ whenever $x_1+x_2 \ge 1$, and ii)  the  implementation of baryon sum rules (see Eq. (6)  in sect. S1 of the SM),  through the choice of a  suitable {factorized} DPD expression} \cite{gauntevo,Golec-Biernat:2014bva}    (see Sect. S3 in SM for details). {In  the present analysis we used the proton PDFs}
from the parametrization of {Refs. \cite{EuropeanMuon:1987obv,SCHAFER1988175} {{referred to as}} EMPDFs in what follows. As already mentioned, for now we adopt  nucleon DPDs,  without   corrections included in modern nucleon SFs, e.g., see \cite{Accardi:2016qay}.}
{Moreover, use has been made of the PDFs evaluated within the LFHM of Ref.  \cite{Kim:2022lng} (see Sect. S5 in SM for details).}  

\begin{figure}[t]
\includegraphics[width=8cm]{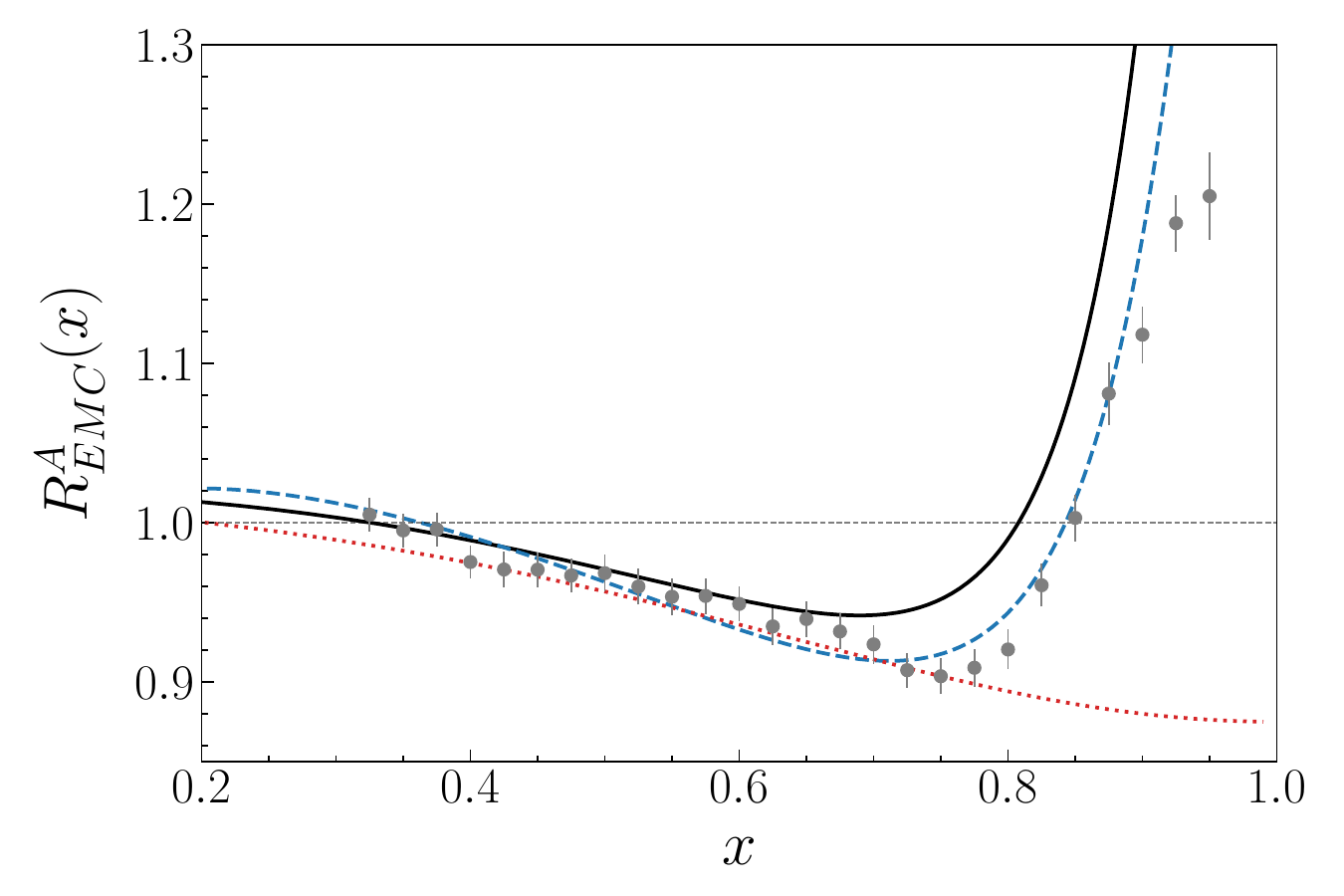}
\includegraphics[width =8cm]{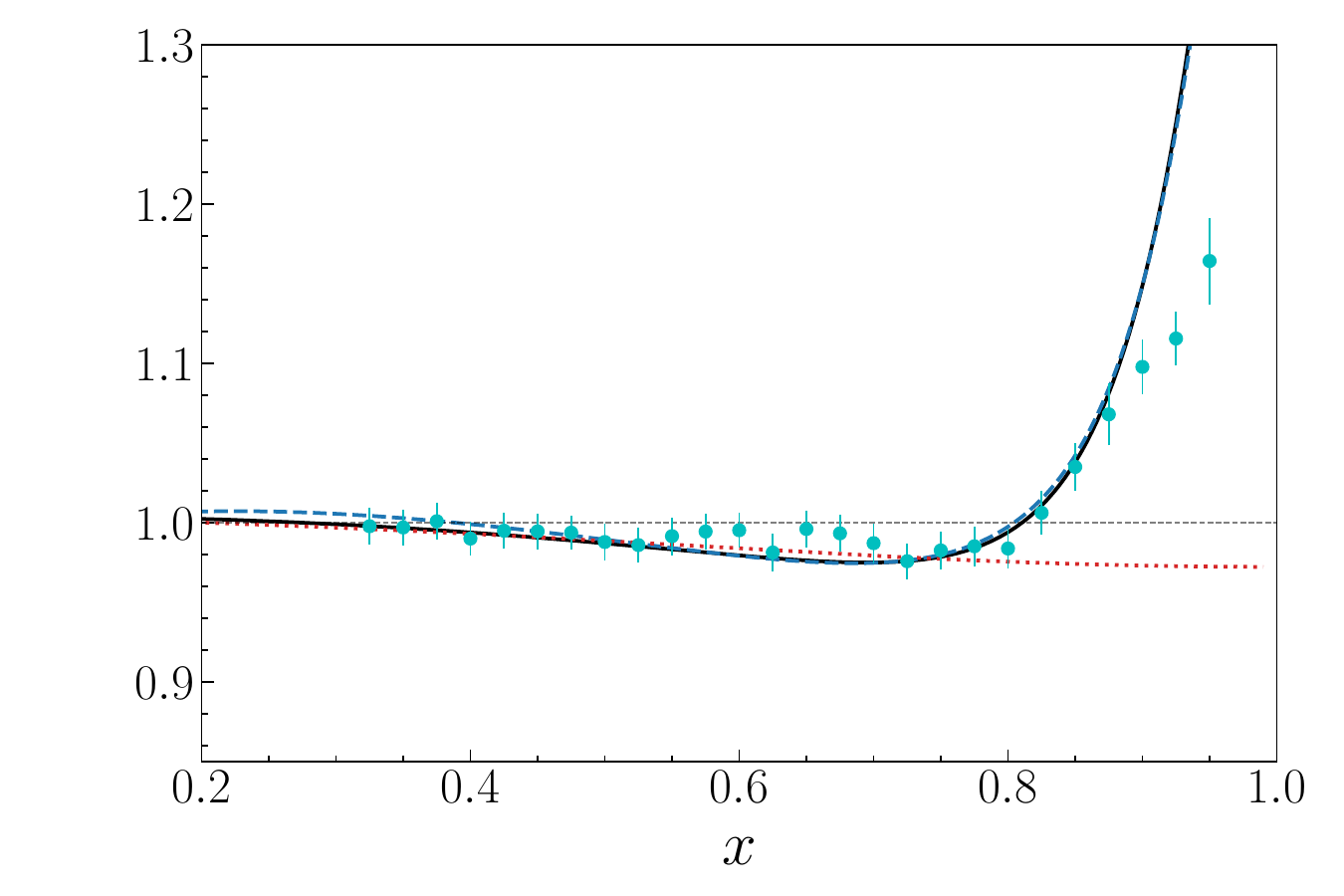}
\caption{ Left panel: The EMC ratio, Eq. (\ref{Eq:EMC}), for $^4$He.  Solid line: nuclear SFs computed from the RHC model \cite{faccioli,noij1}. Dashed line: EMPDF parametrization of the nucleon SFs  \cite{EuropeanMuon:1987obv} with off-shell effects. Dotted line: SFs from the LFHM of Ref. \cite{Kim:2022lng}.  Right Panel: The same as the left panel, but for $^3$He target (notice that  solid and dashed lines largely overlap). Data from Ref. \cite{Seely:2009gt} and reanalyzed in Ref. \cite{Kulagin:2010gd}. 
{The LCMDs {have been} obtained from the {nuclear} wave functions of Refs. \cite{Kievsky:1994mxj,Kievsky:1995uk}.}
}
\label{fig:1}
\end{figure} 

{\em The fully correlated} proton {DPD is given by}
  the relativistic version of  the hyper-central  model (RHC)   (see Refs.
\cite{faccioli,noij1,noij2,noiprl}).  Importantly, the proton DPD, {evaluated within this} model, does not have  free parameters (see  Sect. S4 in  SM) and automatically fulfills both correct support and  GS sum rules.  We adopted this model since it has been already used for investigating to what extent DPCs could be observed in  $ssWW$ cross-sections \cite{noiprl}.

Two different approaches have been implemented for evaluating the nuclear DPDs of light nuclei and for addressing a general pattern involving also the heavy ones. For light nuclei, $2\le A\le 4$, realistic LCMDs are available, Eq. \eqref{Eq:DPD},
while for heavy nuclei the approach  based on  the LFHM of Ref. \cite{Kim:2022lng}, which successfully describes  the EMC ratios of several nuclei in the valence region, has been adopted (see Sect. S5 in  SM for details). {The EMC results {with realistic LCMDs} for heavier nuclei will be presented and discussed in detail elsewhere, while here we are focusing on light nuclei and on an overall pattern for heavier nuclei (see Fig. \ref{fig:2.2} below)}. 

 Our primary aim is to  compare predictions for nuclear DPDs generated by {nucleon structure} models that lead to  similar EMC ratios, {in the valence region $0.3 \leq x \leq 0.7$}. {  In order  to accurately reproduce the EMC ratio within the EMPDF parametrization, additional effects, such as off-shell corrections (OS) are needed.}
 We basically {adopted} the {OS}  of Ref. \cite{Kulagin:2004ie} {(see Sect. S7 in SM)} and proposed some possible extensions for the nucleon DPDs, that we call  "OS1" and "OS2", respectively.  
In the  OS1 case, the OS  {are} applied to the overall  product of the PDFs, while in the OS2 {case}  the OS {are} applied to {each PDF} (for more details see Sect. S7 in SM ).
On the contrary the relativistic version of the fully correlated CQM  gives a reasonable description of the EMC effect for $^3$He and $^4$He (see Fig. \ref{fig:1}), without off-shell corrections.

\section{Results}

In the left panel of  Fig. \ref{fig:1},  the
{ $^4$He } EMC ratio (cf. Eqs.   \eqref{Eq:EMC} and  \eqref{Eq:SF})  is shown. The available data are from Refs.  \cite{Seely:2009gt,Kulagin:2010gd}, and the theoretical lines {make use in Eq. \eqref{Eq:npdf} of the LCMDs  obtained from the wave functions of Refs. \cite{Kievsky:1994mxj,Kievsky:1995uk} and} correspond to: 
i)  the calculation with the RHC \cite{Rinaldi:2018slz,noij1,noij2} model; 
ii) {the evaluation} with the nucleon {EMPDF} PDFs  {and} off-shell effects;  iii)    the outcomes of the LFHM approach \cite{Kim:2022lng} {adapted to $^4$He}, just for the sake of comparison. 
The right panel of Fig. \ref{fig:1}  displays the $^3$He EMC ratio. 
  The {{different models}}  predict  substantially equivalent  EMC effects in the valence region $0.3 \leq x \leq 0.7$ {for both nuclei}.
  Such a result reflects the difficulty of identifying  a conclusive interpretation of the EMC effect. It is worth emphasizing that analogous results can be obtained by using modern nucleon SFs (see, e.g., Refs. \cite{Accardi:2016qay}), but we postpone to a more broad work {{an analysis which takes into account}} {TMC} and higher-twists in both nucleon SFs and DPDs.
 
\begin{figure*}[t]
\includegraphics[width=8cm]{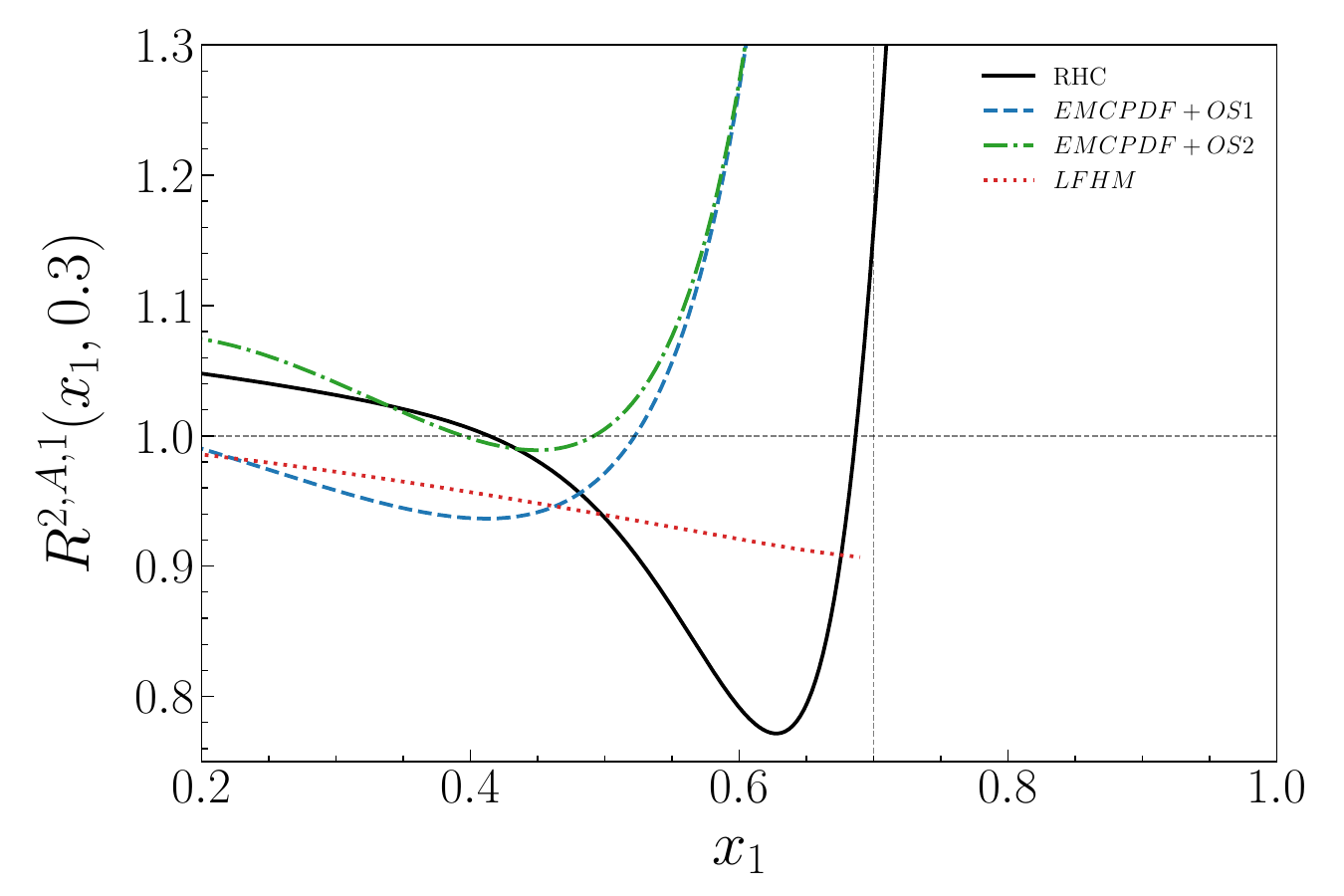}
\includegraphics[width=8cm]{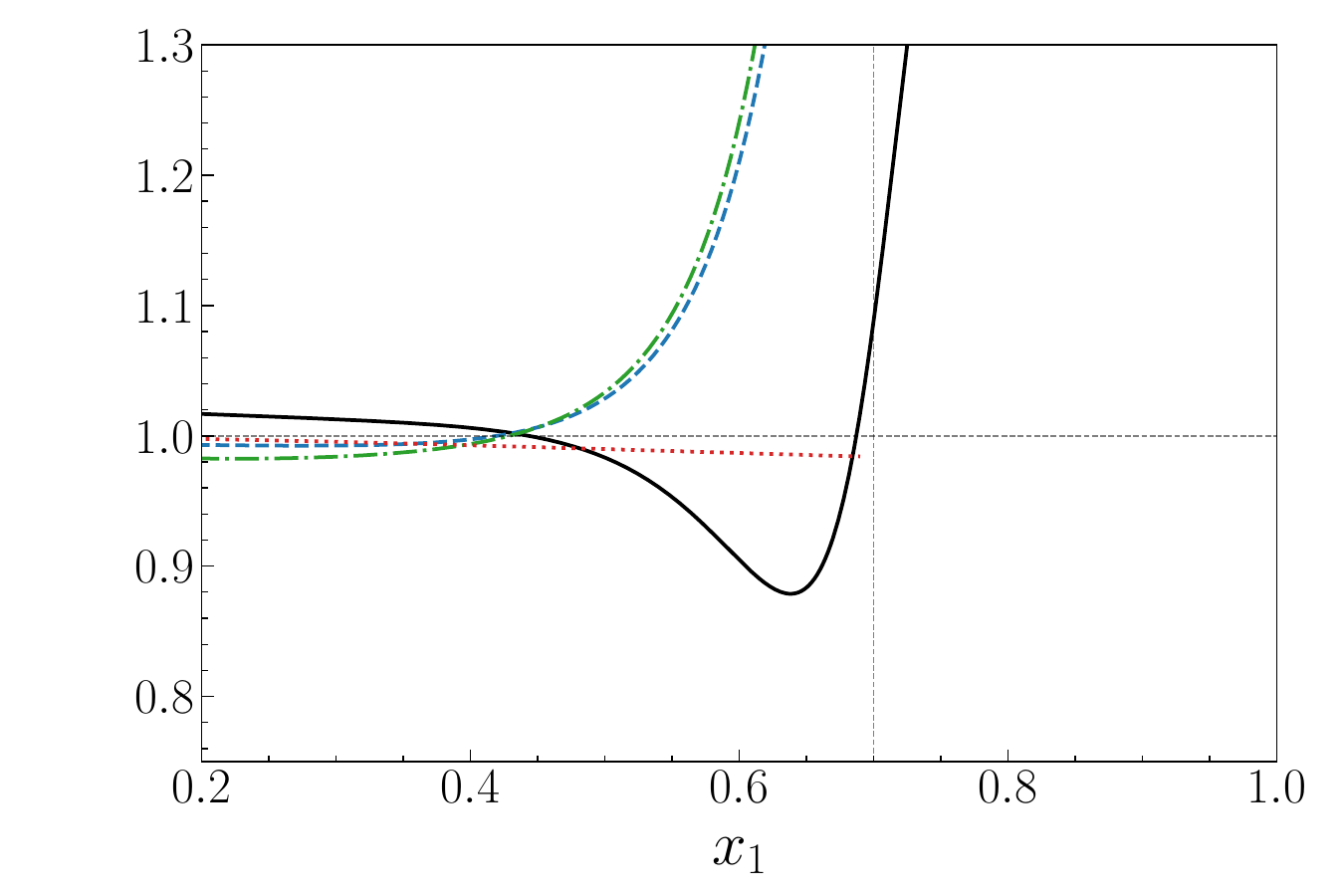}
\caption{ 
Left panel: The ratio Eq. (\ref{Eq:2EMCb}) for $^4$He  with $x_2=0.3$.  Solid line: {results with} proton $(u_vu_v)$-DPD  computed from the RHC model. Dashed line: {calculations with} factorized proton $(u_vu_v)$-DPD computed from the   EMPDF parametrization of Ref.  \cite{EuropeanMuon:1987obv} with  OS1. Dot-dashed line:  the same as the dashed line, but with  OS2. Dotted line: factorized proton  {{$(u_vu_v)$-DPD}} computed from the PDFs of the LFHM  in Ref. \cite{Kim:2022lng}. Right panel: the same as the left panel, but for $^3$He target.}
\label{fig:2}
\end{figure*}

\begin{figure*}[t]
\includegraphics[width=8cm]{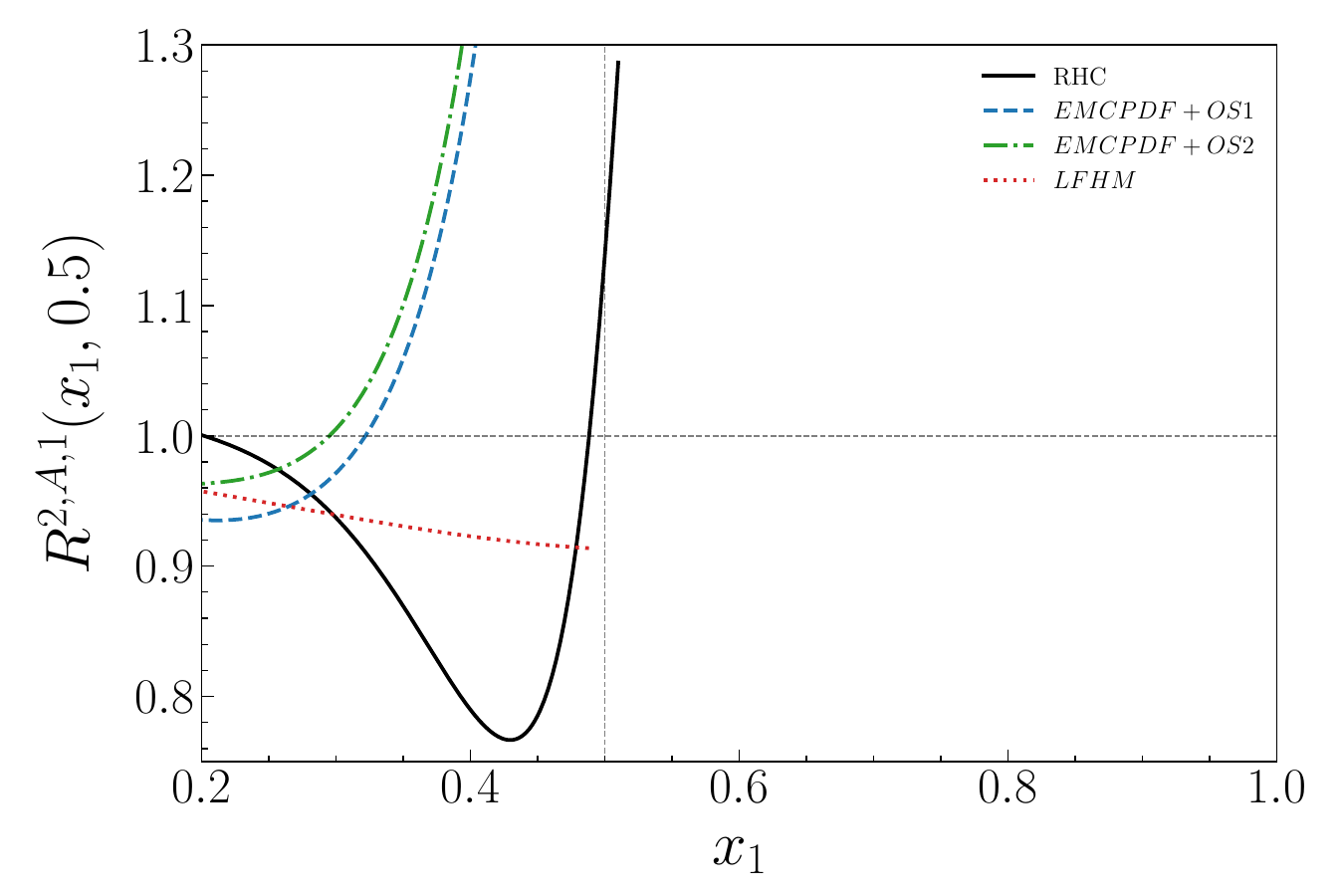}
\includegraphics[width=8cm]{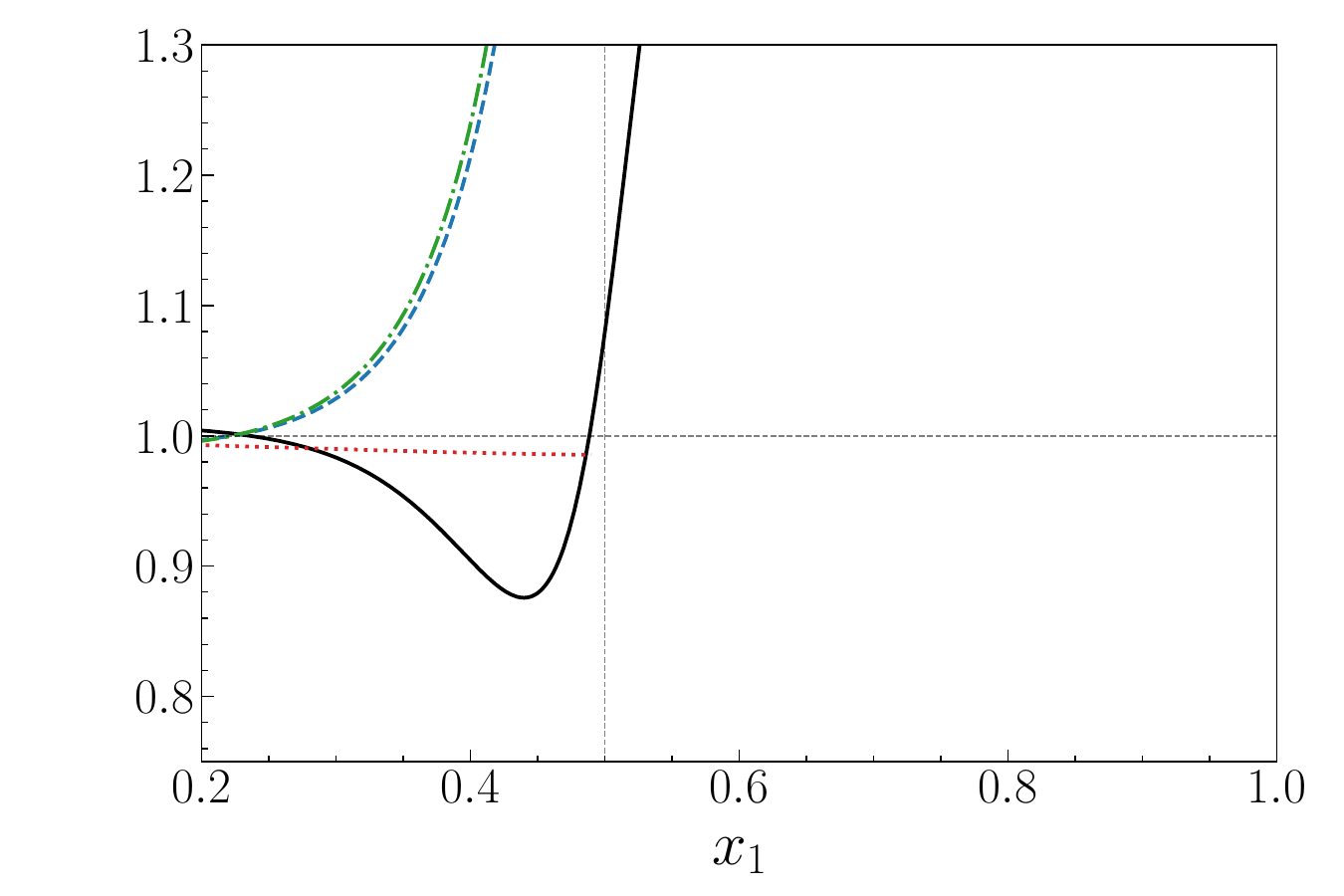}
\caption{ {The same as in  Fig \ref{fig:2}, but for $x_2 = 0.5$}
}
\label{fig:22}
\end{figure*}
  In order to study and pin down partonic effects in  nuclear medium  discovered as a whole by the European Muon Collaboration \cite{EuropeanMuon:1983wih}, we 
 evaluated   the double-EMC ratio, Eq. \eqref{Eq:2EMCa}.
In the case of  factorized DPD models, Eq. \eqref{Eq:2EMCa} exactly reduces to the ratio of the nuclear DPDs, Eq.  \eqref{Eq:DPD}, evaluated at $\textbf{k}_\perp=0$:
\begin{align}
      \label{Eq:2EMCb}
{R^{2,A,1}(z_1,z_2,x_1,x_2) \stackrel{factorized}{\xrightarrow{\hspace*{9mm}} }R^{2,A,1}(x_1,x_2)  = \dfrac{  D^{A,1}_{u_v u_v}(x_1,x_2,0)/
   Z 
    }{ D^{^2H,1}_{u_v,u_v}(x_1,x_2,0) 
 }}.
\end{align}
Notably, also the dependence on the incoming proton $\{z_1,~z_2\}$ cancels out.
In principle, Eq. \eqref{Eq:2EMCb} does not formally hold for the RHC model due to the presence of correlations.  {{To check its validity}}, we numerically analyzed the following ratio (see  {Sect. 6} in SM for details):
 \begin{align}
    {\cal R}_{A2}\left(x_1, x_2, {\bf k}_{\perp}\right)= \frac{D^{A,1}_{u_v u_v}\left(x_1, x_2, {\bf k}_{\perp}\right)}{D^{^2 H,1}_{u_v u_v}\left( x_1, x_2, {\bf k}_{\perp}\right)} ~~ \frac{D^{^2 H,1}_{u_v u_v}\left(x_1, x_2, {\bf 0}_{\perp}\right)}{D^{A,1}_{u_v u_v}\left( x_1, x_2, {\bf 0}_{\perp}\right)} ~.
\end{align}
If ${\cal R}_{A2}\left(x_1, x_2, {\bf k}_{\perp}\right)= 1$, as occurs for the factorised case, then Eq. \eqref{Eq:2EMCa} reduces to Eq. \eqref{Eq:2EMCb}.  
We {{found}} for $^4$He that within a 10$\%$ accuracy ${\cal R}_{A2}$ can be approximated by 1 up to $k_\perp = 3$ GeV in the region  $x_1 + x_2 < 0.95$ . Since  the $k_\perp$ Gaussian fall-off (routinely assumed for TMDs) of both sea and valence nucleon DPDs makes negligible the contribution to the integrals in Eq. \eqref{Eq:2EMCa} from $k_\perp > 3$ GeV, we safely infer that  Eq. \eqref{Eq:2EMCb} is a good approximation also for the RHC DPDs, where correlations are present. Furthermore, one could have an experimental check of the reliability of the approximation
by recalling the relations between $x_i$, $z_i$, $M_W$ and $s_{pN}$.   If  Eq. \eqref{Eq:2EMCb} does not hold, one should observe a residual dependence of $R^{2,A,1}$ on $s_{pN}$, that in the factorized case is exactly cancelled.

Since we are interested in the regions where  $x_1 > 0.2$ and $x_2 > 0.2$, {we considered DPDs for} valence quarks and  evaluated  the double-EMC ratio  for the $(u_v u_v)$-distribution {active in the $W^+W^+$ production.} One should expect {only} a {tiny} dependence on {the typical momentum  scale}  $Q^2\sim M_W^2$, as {it happens} for the usual EMC ratio \cite{Seely:2009gt}. 
Therefore, the study of the residual $Q^2$  effect is deferred to future analyses. 

 The ratios in Eq. (\ref{Eq:2EMCb}),  evaluated {for i) the RHC model and ii)  the factorized DPDs from  LFHM and EMPDF (with OS1 and OS2 modifications) models},  are  shown {{at fixed} $x_2=0.3$  and $x_2=0.5$} in Figs. \ref{fig:2} and \ref{fig:22}, respectively. Remarkably, the different models lead to sizably different results {in clear contrast to what happens in standard EMC ratios}.
This finding suggests that future measurements of nuclear DPDs, {once the DPS2 contribution is under control,} could provide a crucial test for  models of SFs, {based on different mechanisms, but producing almost identical EMC ratios}. {Thus,} such data could potentially clarify the underlying {partonic} dynamics of the EMC effect.
In addition, one should notice that
 for $^3$He the dependence on the choice of the OS model is rather weak (see the right panels in Figs. 2 and 3) while, for the $^4$He nucleus, {one can clearly distinguish} the double-EMC ratios predicted from the OS1 and OS2 models since the higher density of $^4$He results into a greater sensitivity to OS effects. Therefore, one can {argue} that data from DPS processes could shed some light on the possible  {off-shell} mechanism itself in the DPD.

To get {a preliminary} overview  of the $A$-dependence {of the double EMC ratio} beyond the light-nuclei region, we have calculated  the ratio of the nuclear $u-$flavor PDFs in  a nucleus $A$ and {the one} in $^2$H,
{defined by \begin{align}
    R^A_{u}(x) &= \dfrac{  d_u^{A}(x)/
  Z   }{ d_u^{^2H}(x) } ~,
      \label{Eq:EMCu}
   \end{align}}
   as well as  the double-EMC ratio, Eq. \eqref{Eq:2EMCb}. {The calculations have been carried out}  in a wide range of $A$ by using {a factorized  nucleon $(u_vu_v)$-DPD  obtained from} the LFHM.  {Then, we have}  performed the derivative {of $R^A_{u}(x)$}
   with respect to $x$   and the gradient of the double EMC ratio with respect to $\{x_1,x_2\}$ {, namely} $|\vec \nabla R^{2,A,1}(x_1,x_2)|$}. 
{Figure \ref{fig:2.2} shows: }i) for the PDF case,  the quantity $-d R^A_{u}(x)/dx$, evaluated for $x=0.3$ (solid circles), and  ii)  the  double-EMC slope 
$|\vec \nabla R^{2,A,1}(x_1,x_2)|$,
 calculated  for $x_2=0.5$ and $x_1=0.3$.
 The {error bars} {of  $|\vec \nabla R^{2,A,1}(x_1,x_2)|$} in Fig. \ref{fig:2.2} come from the uncertainties in the fits {of the PDF} able  to reproduce the EMC effect data \cite{Kim:2022lng}.
 A similar pattern occurs for different {pairs} $\{x_1, x_2\}$.
The $A$ dependence appears to be more pronounced in the double-EMC ratio compared to the standard EMC effect, hence suggesting a greater sensitivity to the details of the nuclear structure.

\section{Conclusions}
In this study, after recalling the definition of nuclear double-parton distributions (see Ref. \cite{Ceccopieri:2025edn} for more details), we introduced  the definition of a novel quantity: the double-EMC ratio. It is  built by using semi-inclusive $pA$ cross-sections, where only the  double-parton scattering, involving  partons belonging to the same parent nucleon inside the target nucleus, is retained (future studies will address the  theoretical estimate of the other contribution   generated by partons belonging to distinct parent nucleons). {We calculated the double-EMC ratio} by adopting: i) factorized and correlated double-parton distributions {obtained from nucleon-structure} models that predict {almost overlapping} EMC ratios {in the valence region}, ii) realistic light-cone  momentum distributions {{obtained within}} a Poincar\'e-covariant formalism  for $A \leq 4$ and light-front holographic model for $A \geq 2$. Our results {indicate} that  the new quantity is more sensitive to the bound-nucleon structure than the conventional EMC ratio, being less inclusive. Encouraged by the comparisons in Figs. \ref{fig:1}, \ref{fig:2} and \ref{fig:22}   we could argue that future investigations of nuclear double-parton scattering processes have the potential to offer crucial new insights into the origin of the EMC effect, and    pave the  way for an innovative approach for addressing open questions in  both hadron and nuclear physics.

\begin{figure}[t]
\begin{center}
\includegraphics[width=8.7cm]{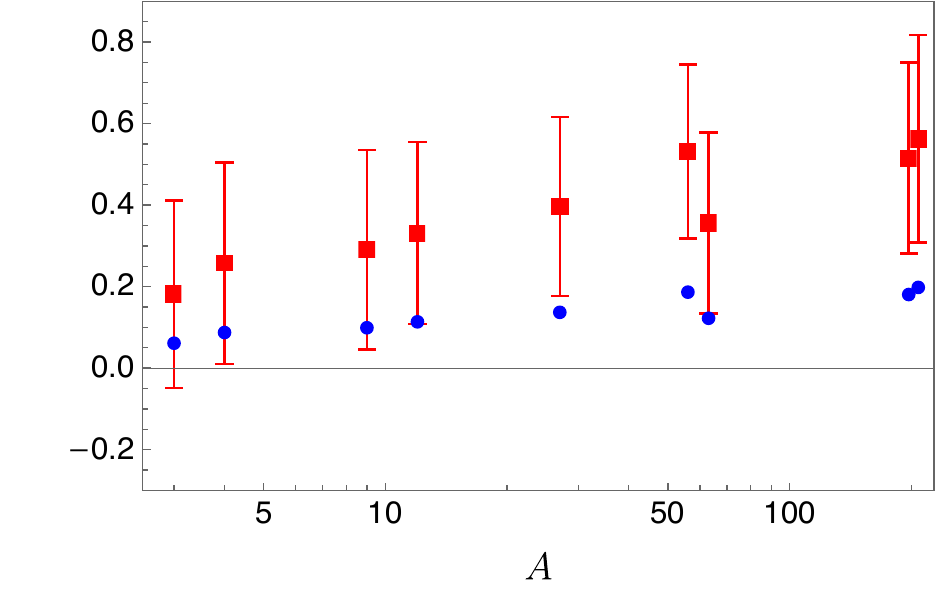}
\caption{ (Color online).
 Calculations of $-d R^A_{u}(x)/dx$ evaluated for $x=0.3$ (lower solid circles) and  $| \vec \nabla R^{2,A,1}(x_1,x_2)|$ calculated  for $x_1=0.3$ and $x_2=0.5$ (full squares) as functions of $A$. {Results obtained}
 within the LFHM  \cite{Kim:2022lng}.
}
\label{fig:2.2}
\end{center}
\end{figure}

\newpage

\clearpage
\appendix
\section*{}

\setcounter{equation}{0}
\renewcommand{\theequation}{S\arabic{equation}}

\setcounter{figure}{0}
\renewcommand{\thefigure}{\arabic{figure}}

\setcounter{table}{0}
\renewcommand{\thetable}{\arabic{table}}

\setcounter{section}{0}
\renewcommand{\thesection}{S\arabic{section}}

\centerline{\bf SUPPLEMENTAL MATERIAL}

\section{Nuclear DPD sum rules} 
{The sum rules (SRs) for nuclear DPDs,  presented in Ref. \cite{Ceccopieri:2025edn},} are generalization to the nuclear case of the Gaunt-Stirling (GS)  sum rules  for hadrons \cite{gauntevo} and are satisfied once  both DPS1 and DPS2 mechanisms  are taken into account. {\it Partial sum rules} (PSRs) for {both} the DPS1 and DPS2 mechanisms  were {also} introduced. In this Section, we summarize some of the results discussed in Ref. \cite{Ceccopieri:2025edn}, as well as numerical results for the SRs with focus on  the DPS1 mechanism.

{Let us first}
{recall the expression for the nuclear DPD corresponding to the DPS1 mechanism (see Eq. (1) of the main text):
\begin{equation}
D_{i j}^{A,1}\left(x_{1}, x_{2}, k_{\perp}=0\right)=\sum_{\tau=p, n} \int_{\xi_{min}}^{1} d \xi \frac{\bar{\xi}^2}{\xi^2} \bar \rho^A_{\tau}(\xi) D_{i j}^{\tau}\left(\frac{x_{1}}{\xi} \bar{\xi}, \frac{x_{2}}{\xi} \bar{\xi}, 0\right),
\label{Nuclear_DPDs}
\end{equation}
 where: $i)$  $\bar \xi = m/M_A$, {$ii)$ $x_1+x_2 \leq M_A/m$, $iii)$ $\bar \rho$ is the one-body light cone nuclear density normalized to $A$ and $iv)~ \xi_{min} = (x_1+x_2)\bar \xi$ }. In particular, in Ref. \cite{Ceccopieri:2025edn} it has been proven the nuclear number sum rule:
 \begin{align}
    \int_{0}^{1/\bar{\xi}-x_{1}} d x_{2} D_{i j_{{v}}}^{A,1}\left(x_{1}, x_{2}, 0\right) =   \sum_{\tau=p, n} \begin{cases}\left(N^\tau_{j_{{v}}}-1\right) d_{i}^{A,\tau}(x_1) & i=j \\
N_{j_{{v}}}^{\tau} d_{i}^{A,\tau}(x_1) & i \neq j
\\
\left(N^\tau_{j_{{v}}}+1\right) d_{i}^{A,\tau}(x_1) & i=\bar j
\end{cases},
\label{NSR_DPS0}
\end{align}
where {$N^\tau_{j_{{v}}}$} is the number of valence quark of flavor $j$ in the $\tau$-th nucleon, {and}  $d_{i}^{A,\tau}(x)$  the nuclear PDF:

 \begin{align}
d^{A,\tau}_i(x) =
\int_{ {\xi_{m}}}^{1} d \xi~   \bar \rho^A_{\tau} (\xi) \dfrac{\bar \xi}{\xi} d^\tau_i \left(x \dfrac{\bar \xi}{\xi}\right) ~, 
\label{dAti}
\end{align}
{{with} $\xi_{m}=x~\bar\xi$}.
 From the above relation one can obtain the {{\em partial}} normalization of the nuclear DPD corresponding to the DPS1 mechanism:
\begin{align}
   \int_0^{1/\bar \xi} dx_1~ \int_{0}^{1/\bar{\xi}-x_{1}} d x_{2} D_{i j_{{v}}}^{A,1}\left(x_{1}, x_{2}, 0\right) =   \sum_{\tau=p, n} \begin{cases}\left(N^\tau_{j_{{v}}}-1\right)N^\tau_{i_v}  & i=j \\
N_{j_{{v}}}^{\tau} N^\tau_{i_v} & i \neq j\end{cases},
\label{NSR_DPS1}
\end{align}
{We remark that the above condition is valid only for the low energy-scale, where the only non-zero distribution is the valence one \cite{Ceccopieri:2025edn}}.
{This result is particularly {valuable}  {{to give an overall estimate of}} the relative {contribution} of the DPS1 mechanism to the total nuclear DPD. {{To this end one can perform a comparison with}} the {{{\em global}} normalization of the DPD of the $A$ nucleus:}
  \begin{align}
   \int_0^{1/\bar \xi} dx_1~ \int_{0}^{1/\bar{\xi}-x_{1}} d x_{2} D_{i j_{{v}}}^{A}\left(x_{1}, x_{2}, 0\right) =   \begin{cases}\left(N^A_{j_{{v}}}-1\right)N^A_{i_v}  & i=j \\
N_{j_{{v}}}^{A} N^A_{i_v} & i \neq j\end{cases},
\label{NSR_DPS}
\end{align}
where now $N^A_{i_v(j_v)}$ is the total number of valence quarks of flavor $i(j)$ in the nucleus.
{The above expression is a generalization to a nuclear target of the baryon sum rule developed for the proton in Ref. \cite{gauntevo}:}

\begin{align}
    \int_0^{1} dx_1~ \int_{0}^{1-x_{1}} d x_{2} D_{i j_{{v}}}^{p}\left(x_{1}, x_{2}, 0\right) =   \begin{cases}\left(N^p_{j_{{v}}}-1\right)N^p_{i_v}  & i=j \\
N_{j_{{v}}}^{p} N^p_{i_v} & i \neq j\end{cases},
\label{Eq:baryon}
\end{align}

Theoretical results for the right-hand side of Eqs. \eqref{NSR_DPS1} and \eqref{NSR_DPS}
 are shown in Tab. \ref{table:1}.
 From a numerical point of view, the left-hand side of Eq. (\ref{NSR_DPS1}) has been calculated {for the $^2$H, $^3$He and  $^4$He nuclei}
 using the  nucleon DPD corresponding to the  {RHC model (see Sect. S4 for some details)}}, {which fulfills analytically the 
 GS sum rules \cite{gauntevo} for the nucleon,}} {{and}} proper 
 LCMD of the $^2$H, $^3$He and  $^4$He nuclei \cite{Pace:2001vg,DelDotto:2016vkh,Pace:2020ned,Alessandro:2021cbg,Pace:2022qoj}.
{Numerical} results are consistent with the theoretical expectations {at 0.001 $\%$ level for $^2$H  and 0.01 $\%$ level for $^3$He, $^3$H and $^4$He. } Further details can be found in Ref. \cite{Ceccopieri:2025edn}.
 As one can see {in Table \ref{table:1}}, depending on the kind of nucleus and on the combination of the quark flavors, the DPS1 contribution {cannot be disregarded}.  The relative weights range { about $30\%$ to $50\%$, and $10\%$ to $20\%$}, for the set of light nuclei and flavor combinations we have considered.

\begin{table}[h]
\begin{center}
\scalebox{1}{
\begin{tabular}{|c c|c|c|c|c|}
\hline
\hline
 & &  $^2$H  &$^3$H & $^3$He & $^4$He \\
 
\hline
\multirow{2}{*}{uu}

& \mc{1}{|l|}{SR}  & 6 &  20 & 20 & 30   \\

& \mc{1}{|l|}{PSR}  & 2  & 2 & 4 &  4 \\  

\hline

\multirow{2}{*}{ud}

& \mc{1}{|l|}{SR}  & 9  &  12 & 12 & 36   \\

& \mc{1}{|l|}{PSR}  & 4 & 6 & 6 & 8  \\  

\hline

\multirow{2}{*}{dd}

& \mc{1}{|l|}{SR}  & 6  &  20 & 20 & 30    \\

& \mc{1}{|l|}{PSR}  & 2 & 4 & 2 & 4  \\ 

\hline
     
\hline
\end{tabular}
}
\end{center}
\caption{Theoretical   results for both SRs (rhs of Eq. (\ref{NSR_DPS})) and {PSRs} (rhs of Eq. (\ref{NSR_DPS1})), for different light nuclei and different flavors. All the values in the Table were numerically checked {(see text)}.}
\label{table:1}
\end{table}

\section{kinematics of electroweak-bosons production in $pA$ collisions. }

This Section is devoted to the high-energy collisions of a proton and a nucleus with atomic number $A$ {{which}} produce two same-sign $W$'s: 
\begin{equation}
A(p_A) + p(p_{in}) \rightarrow W^+(q_1) + W^+(q_2) +X
\end{equation}
We consider the {target} nucleus moving along $+\hat{\bf e}_z$ and the {incoming} proton moving along $-\hat{\bf e}_z$,  in the  centre of mass frame of the proton, with 4-momentum $p_{in}$, and of a {bound} nucleon $N$  with 4-momentum $p_N$. Then {the square of }the per-nucleon centre-of-mass energy of the collision is 
given by $(p_{in}+p_N)^2=s_{p_N}$.
With those assumptions the momentum {(Cartesian and light-cone)}  components are given by {(in the ultra-relativistic approximation: $p^0_{in}=p^0_N=|{\bf p}_z|$, being $s_{p_N}\gg M^2_N$)}:
\begin{eqnarray}
    p_N=\frac{\sqrt{s_{p_N}}}{2} (1,\mathbf{0}_\perp,1), &&\hspace{1cm} 
    p^+_N=\sqrt{s_{p_N}/2} , \quad p^-_N=0,\\
    p_{in}=\frac{\sqrt{s_{p_N}}}{2} (1,\mathbf{0}_\perp,-1), &&\hspace{1cm} p_{in}^-=\sqrt{s_{p_N}/2} , \quad p^+_{in}=0,
\end{eqnarray}
{with $a^\pm_i=(a^0_i \pm a^3_i)/\sqrt{2} $}.
In the framework of DPS, the two $W$'s are produced by the annihilation 
of two pair of partons: {one pair 
from the {nucleon inside the} nucleus, {with momenta $p_1$, $p_2$ and longitudinal momentum fractions $x_1$, $x_2$, and the other from the incoming proton, with momenta $p_3$, $p_4$ and longitudinal momentum fractions $z_1$, $z_2$}}. {In the above described frame, one writes}:
\begin{eqnarray}
    p_1 = x_1 \frac{\sqrt{s_{p_N}}}{2} (1,\mathbf{0}_\perp,1)~, \hspace{2cm}
    p_3 = z_1 \frac{\sqrt{s_{p_N}}}{2} (1,\mathbf{0}_\perp,-1)~, \\ 
    p_2 = x_2 \frac{\sqrt{s_{p_N}}}{2} (1,\mathbf{0}_\perp,1)~, \hspace{2cm} 
    p_4 = z_2 \frac{\sqrt{s_{p_N}}}{2} (1,\mathbf{0}_\perp,-1)~, 
\end{eqnarray}
where $x_{1(2)}=p^+_{1(2)}/p^+_N$,  $z_{1(2)}=p^-_{3(4)}/p^-_{in}$. {Then, the same-sign W's production reads}: 
\begin{eqnarray}
    i(p_1)+j(p_3) \rightarrow W(q_1);  \\
    k(p_2)+l(p_4) \rightarrow W(q_2) ~, 
\end{eqnarray}
where $i,j,k,l$ indicate the flavor combination relevant for the process:
$u\bar{d}$ annihilation is relevant for $W^+$ production, $d\bar{u}$ for $W^-$ and finally $u\bar{u}+d\bar{d}$ for Drell-Yan pairs, with electroweak charges and Cabibbo-suppressed channels  omitted for clarity (see, e.g.  Ref. \cite{noiprl} for further details).
Given the above kinematic, both annihilations must give an object whose invariant mass is:
\begin{eqnarray}
    (p_1+p_3)=q_1^2=M_W^2; \\
    (p_2+p_4)=q_2^2=M_W^2. 
\end{eqnarray}
From those constraints, one immediately derives two conditions on the product of momentum fractions for each scattering {(recall $p^2_i\sim 0$)}: 
\begin{eqnarray}
    x_1 z_1 =M_W^2/s_{pN}~, \qquad \qquad
    x_2 z_2=M_W^2/s_{pN}.  
\end{eqnarray}
which is fixed once the boson mass and the center of mass energy of the collisions are set. 
Likewise, let us define the rapidity of a particle in a given frame:
\begin{equation}
    Y=\frac{1}{2} \ln \frac{E+p_z}{E-p_z}
\end{equation}

{{The momenta of the two vector bosons read}}:
\begin{eqnarray}
    q_1= \frac{\sqrt{s_{pN}}}{2} (x_1+z_1,\mathbf{0}_\perp,x_1-z_1); \\ 
    q_2= \frac{\sqrt{s_{pN}}}{2} (x_2+z_2,\mathbf{0}_\perp,x_2-z_2); 
\end{eqnarray}
By inserting  {{these components}} in the rapidity definition, one immediately finds {the rapidities of the two bosons}:
\begin{eqnarray}
    Y_1=\frac{1}{2} \ln \frac{x_1}{z_1}~, \qquad \qquad
    Y_2=\frac{1}{2} \ln \frac{x_2}{z_2} 
\end{eqnarray}
We may catalog the $WW$ phase space according to their rapidities making some assumptions on the momentum fractions:
\begin{itemize}
    \item $x_1 \ll z_1$ and $x_2 \ll z_2$: both $WW$  are produced along the proton  moving direction ($- \hat{\textbf{e}}_z$) with large and negative rapidities.
    \item $x_1 \gg z_1$ and $x_2 \gg z_1$: both $WW$ are produced along the nucleus  moving direction ($+\hat{\textbf{e}}_z$) with large and positive rapidities.
    \item $x_1 \ll z_1$ and $x_2 \gg z_2$ or $x_1 \gg z_1$ and $x_2 \ll z_2$: the $WW$ are produced in opposite hemisphere.
\end{itemize} 
Recalling that valence quarks distributions prevail for $x_i \to 1$, while the sea distributions dominate at $z_i \to 0$, one good candidate for the proposed analysis is $D^A_{u_v u_v}(x_1,x_2)$, that can be singled out by choosing the region $x_i\gg z_i$.  
Therefore slicing in rapidity proves to be helpful in selecting
desired ranges in momentum fractions.
We provide here additional constraints relevant for a DPS scattering:
\begin{eqnarray}
    0 \le z_1+z_2 \le 1 &:& \mbox{proton double-parton distribution}; \nonumber\\
    0 \le x_1+x_2 \le 1 &:& \mbox{nuclear double-parton distribution in DPS1}; \nonumber\\
     0 \le x_1+x_2 \le 2 &:& \mbox{nuclear single parton distributions in DPS2}; \nonumber
\end{eqnarray}
One should notice that the second inequality is exact only when $\bar \rho^A(\xi) \propto \delta(\xi - \bar \xi)$. However, as follows from calculations on light-nuclei as in Ref. \cite{Ceccopieri:2025edn}, the distribution $D_{u_v u_v}^{A,1}(x_1,x_2)$ is sizeable only for $x_1 + x_2 \leq 1$. Similar arguments can be applied also to the third inequality, recalling that in the DPS2 mechanism two nucleons partecipate in the process. 

\section{  The nucleon factorized DPD\MakeLowercase{s} }

In this Section, we present the approach adopted to build the   DPDs from the product of usual PDFs. Let us remark that this kind of Ansatz is often used for phenomenological analyses, being DPDs {virtually} unknown.
In the most simple case, the DPD is given by the product of PDFs $d_i^\tau(x)$:
\begin{equation}
    D^\tau_{ij}(x_1,x_2,k_\perp) = d_i^\tau(x_1)d_j^\tau(x_2)g(k_\perp).
    \label{Eq:facto}
\end{equation}    
{Plainly}, this approach has some drawbacks: i)  the support is the union of the supports of the PDFs, i.e. $x_1 \le 1$ and $x_2  \le 1$, hence there is a non-physical tail for $x_1+x_2 >1$; ii) the Gaunt-Stirling sum rules are not fulfilled; iii) this quantity is clearly non symmetric under the  exchange of $x_1 \leftrightarrow x_2$ {or} $i \leftrightarrow j$;
iv) there is no correlation between $x_1$ and $x_2$ and
v) the shape of $g(k_\perp)$ is not well constrained. 
{{Since we are mainly interested in ratios, where $g(k_\perp)$ cancels out, the last issue {will be discussed elsewhere and in the following we put $g(k_\perp=0)=1$}}}.   In order to fix the first two issues, there are different ways to proceed. One can add a theta-function to force the correct support $x_1 + x_2 \le 1$ and  can {also implement} the symmetry in $i,j$ and $x_1, x_2$ following Ref.  \cite{Golec-Biernat:2014bva}:
\begin{align}
\nonumber
    D^\tau_{ij}(x_1,x_2,0) = & \frac{\lambda^{\tau}_{i j}}{4}\Bigg[ \frac{1}{1-x_2} \left[ d_i^\tau\left(\frac{x_1}{1-x_2}\right)d_j^\tau(x_2) + d_j^\tau\left(\frac{x_1}{1-x_2}\right)d_i^\tau(x_2) \right]+ 
    \\
    & +
\frac{1}{1-x_1} \left[ d_j^\tau(x_1) d_i^\tau \left(\frac{x_2}{1-x_1}\right) + d_i^\tau(x_1) d_j^\tau\left(\frac{x_2}{1-x_1}\right) \right] \Bigg] \theta(1-x_1 - x_2),
\label{DPDs_cambio_variabile}
\end{align}
with $\lambda^\tau_{i j} = 1$ except for $\lambda^p_{d_v d_v} = \lambda^n_{u_v u_v} =0$. 
With this Ansatz, the correct {baryon} sum rules Eq. \eqref{Eq:baryon}  are fulfilled for the valence quarks and the DPD automatically  vanish when $x_1+x_2$ is approaching one.


\section{The relativistic Hypercentral Constituent Quark Model for double-parton distributions}

In this Section, some details about the    {relativistic hypercentral constituent quark (RHC) model are discussed. This model is applied to obtain the} {fully correlated nucleon   PDF and DPD {{used}} in the evaluation of both single and double EMC ratios.}   In the {adopted} framework,  PDFs and DPDs are evaluated {by using}  the nucleon wave function {in valence approximation} corresponding to a  {hypercentral} potential which parametrizes some features of QCD. 
The {relativistic} {version of the} {hypercentral constituent quark} model, discussed and introduced in \cite{faccioli,Ferraris:1995ui}, has been already used for obtaining DPDs in Ref. \cite{noij1}.  In Refs. \cite{jhepc,Ceccopieri:2017oqe,noiPLB} it was found that this model leads to predictions of the quantity called {\it effective cross-section} {{not far from}}  experimental analyses, within the errors. Such a quantity   is given by the ratio between  the product of {single-parton} cross-sections and the {double-parton}  one, taking in the numerator the same final particles present in DPS under scrutiny. 
Therefore, $\sigma_{eff}$ represents an experimental observable that can be calculated from DPDs.  In particular within the factorized DPDs Ansatz, the effective cross-section reduces to \cite{gauntevo}
\begin{align}
    \sigma_{eff} = \dfrac{1}{\displaystyle \int \dfrac{d^2 k_\perp}{(2 \pi)^2}g(k_\perp)^2 },
    \label{eq:sigma_eff}
\end{align}
 where $g(k_\perp)$ is the effective form factor \cite{Rinaldi:2018slz} appearing in the factorized DPDs Ansatz  (see Sect. S2).
{ In the fully correlated RHC model, one  predicts $\sigma_{eff} \sim 20$ mb  for the same-sign {ss} $WW$ production, that favorably compares with the recently-extracted experimental value $\sigma_{eff} = 12.2^{+2.9}_{-2.2}$ mb \cite{CMS:2022pio}}.

{Since the aim of the present analysis is the evaluation of the double EMC {ratio}, {for gaining information on the standard} EMC  one,   as a first step, we show the results   of  the nuclear PDFs, {that are the main ingredient for the theoretical calculation of the EMC ratio}, (see Eq. (2) of the main text). {In Fig. 1,  we present the $u$-flavor nuclear PDFs within the RHC model. It is worth mentioning that  the  RHC CQM does not have free parameters. }

\begin{figure}[t]
\includegraphics[width=8.7cm]{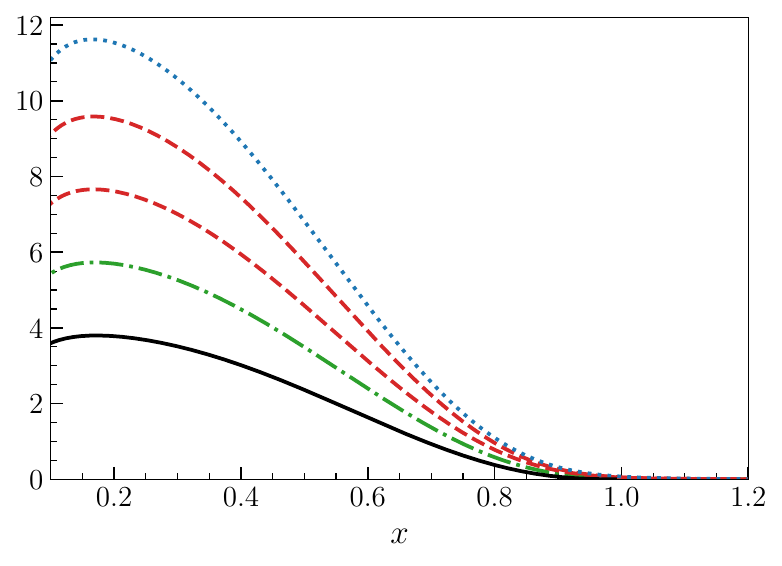} 
\caption{ (Color online).  
The nuclear PDFs for the $u$-flavor evaluated within the  RHC model. Dotted line:   $^4$He. Dashed line: $^3$He. Dot-dashed line:  $^2$H. Solid line: $u$-PDF in the free proton. 
}
\label{fig:1}
\end{figure}

\section{ The light-front holographic model model for the DPDs}
In this Section,  details on the  LFHM  \cite{Kim:2022lng}, used to evaluate nuclear PDFs, SFs and generalized  to calculate nuclear DPDs, will be provided. In this model, the {full} nuclear dynamics is encoded in  two parameters, fixed by fitting EMC data. This approach enables a straightforward extension to DPDs, even for heavy nuclei, where it is challenging to get  reliable wave-functions. However, it is important to mention that this model cannot reproduce the Fermi motion in the EMC ratio.

Starting with the free-nucleon case, we   retain  only  valence quarks and assume the isospin symmetry. Then PDFs  read \cite{Kim:2022lng}:
\begin{equation}
\begin{split}
     &u^p_v(x) = d^n_v(x) = \frac{3}{2}q_3(x) + \frac{1}{2}q_4(x) \\
     & d^p_v(x) = u^n_v(x) = q_4(x)
\end{split}
\end{equation}
where:
\begin{equation}
    q_\alpha(x) = \frac{\Gamma(\alpha-\frac{1}{2})}{\sqrt{\pi}\Gamma(\alpha-1)} \left[ 1 - \omega(x) \right]^{\alpha-2} \omega(x)^{-\frac{1}{2}} \omega'(x),
\end{equation}
with:
\begin{equation}
    \omega(x) = x^{1-x}e^{-a(1 - x)^2},
\end{equation}
and $a = 0.531 \pm 0.037$. The SFs {for the proton ($p$) and the neutron ($n$)}, can be written as follows:
\begin{equation}
\begin{split}
    & F_2^p(x) = \frac{4}{9} x u^p_v(x) + \frac{1}{9} x d^p_v(x) \\
    & F_2^n(x) = \frac{4}{9} x u^n_v(x) + \frac{1}{9} x d^n_v(x)
\end{split}   
\end{equation}

In Ref. \cite{Kim:2022lng},  some modifications of the PDFs {for bound nucleons} {are implemented}. In particular, these changes depend only on the difference $q_4(x) - q_3(x)$ which is weighted  by a parameter $\delta r_\tau^A$ different for each nucleon ($\tau$) and nucleus. The modified PDFs read as follows:
\begin{equation}
    \begin{split}
        & \tilde{u}_v^\tau(x) = u^\tau_v(x) + 3 \delta r_\tau^{ A} \big[ q_4(x) - q_3(x) \big] \\
        & \tilde{d}_\tau^n(x) = d^\tau_v(x) + 3 \delta r_\tau^{ A} \big[ q_4(x) - q_3(x) \big],
    \end{split}
    \label{Eq:PDFmod}
\end{equation}
and then the modified {nucleon} SFs  become:
\begin{equation}
        \tilde{F}_2^\tau(x) = F_2^\tau(x) + \frac{5}{3} x \delta r_\tau^{ A} \big[ q_4(x) - q_3(x) \big].
        \label{Eq:LFHMPDF}
\end{equation}
Therefore, for each nucleus, there are two parameters ($\delta r_P^A$ and $\delta r_N^A$) to be fitted from the data of the corresponding EMC effect. {In conclusion,}  nuclear SFs can be written as follows:
\begin{equation}
    \begin{split}
    F_2^A(x) &= Z \tilde{F}_2^p(x) + (A-Z)\tilde{F}_2^n(x) \\
    & = ZF_2^p(x) + (A-Z)F_2^n(x) + \frac{5}{3}x\big[ q_4(x) - q_3(x)\big] \big[Z \delta r_p^A + (A-Z) \delta r_n^A \big].
    \end{split}
\end{equation}
One should notice that since all distributions appearing in the above equations vanish for $x \to 1$ for each nucleus, {hence} the model fails in reproducing the Fermi motion in the EMC ratio. However, for $x \in \big[ 0.3, 0.7\big]$, 
 the model can describe EMC effect  also for heavy nuclei, {given the presence of adjusted parameters, as mentioned above. }
Therefore, we can generalize this model to evaluate  DPDs. To {this end},
we can use the factorized Ansatz in Eq. \eqref{DPDs_cambio_variabile} to build the bound nucleon DPDs from the  bound PDFs,  obtaining the following nuclear DPDs, corresponding to the DPS1 mechanism:
\be
{D^{A,1}_{ij}(x_1,x_2) = Z \Tilde{D}^{p}_{ij}(x_1,x_2) + (A-Z) \Tilde{D}^{n}_{ij}(x_1,x_2)},
\label{eq:D_A1}\ee
{where $ \Tilde{D}^{p(n)}_{ij}(x_1,x_2)$ is obtained from Eq. \eqref{DPDs_cambio_variabile} by using the PDFs of Eq. (\ref{Eq:PDFmod}).}
{By using the nuclear DPDs in Eq. \eqref{eq:D_A1}, we have investigated the dependence of double-EMC ratio  for several nuclei.} {Recall} that {the LFHM} is quite useful to make predictions for heavy nuclei, for which the LCMDs are not yet available.

The single- and double-EMC like ratios, respectively  in Eqs. {(13) and (15)} of the main text, have been evaluated and shown in Figs. \ref{Fig:RA1} and \ref{Fig:RA2} for the following nuclei: $^4$He (full lines), $^{12}$C (dashed lines), $^{63}$Cu (dotted lines), $^{197}$Au (dot-dashed lines) and $^{208}$Pb (long dashed lines) (notice that the dashed and dotted lines largely {overlap}). As one can see in Figs. \ref{Fig:RA1} and \ref{Fig:RA2}, while the $A$ dependence of the single-EMC ratio is quite small, less than 10$\%$ at $x = 0.7$ {going} from $^4$He to $^{208}$Pb (see the left panel of Fig. \ref{Fig:RA1}),  the variations of the double-EMC ratio   can be  {as high as} 15$\%$ for $x_2=0.2$ (see the right panel of Fig. \ref{Fig:RA1}) and  {as} 20$\%$ for $x_2=0.3$ and $x_2=   0.5$ (see Fig. \ref{Fig:RA2}).
Such an interesting result shows the stronger dependence of $R^{2,A,1}$ on $A$ w.r.t. $R^A_{EMC}$.
In the future, other models {will} be used to validate these  {outcomes}.

\begin{figure}[t]
\includegraphics[width=8.7cm]{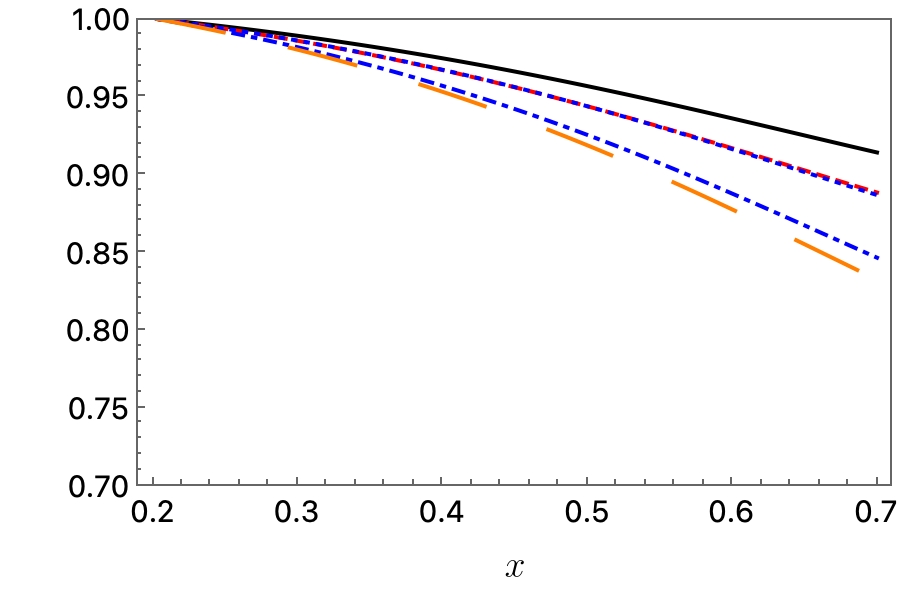} 
\includegraphics[width=8.7cm]{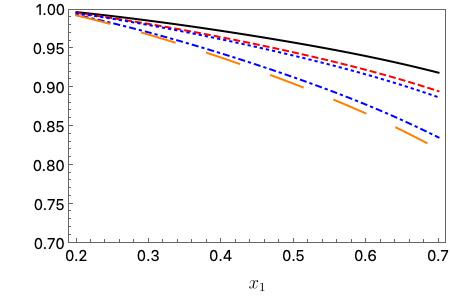} 
\caption{ (Color online).  Left panel: calculation of the single-EMC ratio in Eq. (15) of the main text. Right panel: Calculation of the double-EMC ratio  { $R^{2,A,1}(x_1,0.2)$} in Eq. (13) of the main text. In both panels the calculation for the following nuclei are shown: $^4$He (full lines), $^{12}$C (dashed lines), $^{63}$Cu (dotted lines), $^{197}$Au (dot-dashed lines) and $^{208}$Pb (long dashed lines).
}
\label{Fig:RA1}
\end{figure}

\begin{figure}[t]
\includegraphics[width=8.7cm]{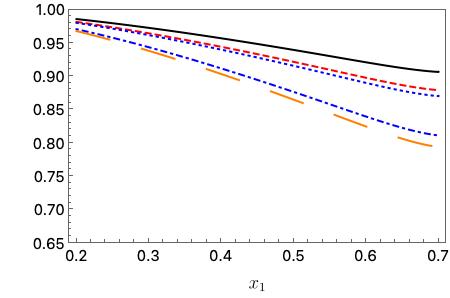} 
\includegraphics[width=8.7cm]{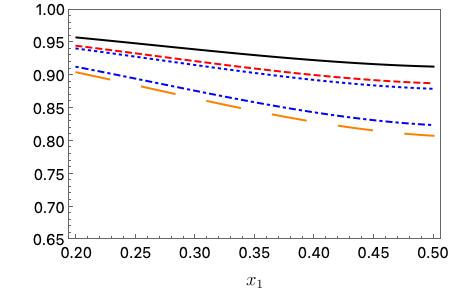} 
\caption{ (Color online). Left panel:  the same as in  the right panel of Fig. \ref{Fig:RA1} for  {$R^{2,A,1}(x_1,0.3)$}. Right panel: the same of the right panel of  Fig. \ref{Fig:RA1} for {$R^{2,A,1}(x_1,0.5)$}.
}
\label{Fig:RA2}
\end{figure}

\section{Violation of the factorized Ansatz in double-EMC ratios}

In this section we investigate to what extent {a DPD fully-correlated  model violates Eq. {{(12)}} of the main text, that formally holds for the DPD factorized Ansatz.}   We remind that the main interest of our analysis is the following ratio  {(cf. {Eq.}} (11) in the main text)}:
\begin{align}
 \label{ratio1}
    R^{2,A,1}(z_1,z_2,x_1,x_2) 
    &= {1 \over Z}~\dfrac{ S^{pA,1}_{\bar d \bar d u_v u_v}(z_1,z_2,x_1,x_2)}{ S^{p ^2H1}_{\bar d \bar d u_v u_v}(z_1,z_2,x_1,x_2)}   
    \\
\label{ratio1k}
    &= {1 \over Z}~\dfrac{\displaystyle \int {d^2 {\bf k}_{\perp}\over (2 \pi)^2} 
 D^p_{\bar d \bar d}\left(z_1, z_2,- {\bf k}_{\perp} \right) D^{A,1}_{u_v u_v}\left( x_1, x_2, {\bf k}_{\perp}\right)   }{\displaystyle  \int {d^2 {\bf k}_{\perp}\over (2 \pi)^2} 
 D^p_{\bar d \bar d}\left(z_1, z_2,- {\bf k}_{\perp} \right) D^{^2 H,1}_{u_v u_v}\left( x_1, x_2, {\bf k}_{\perp}\right)   }.
\end{align}
 One should notice that, for the correlated case, the dependence on $z_1$ and $z_2$ survives, i.e., {there is} a residual  dependence on $(p_{in}+p_N)^2=s_{pN}$ (see the main text before Eq. (7)). {The main issue one meets when  the ratio in Eq. \eqref{ratio1k} is calculated, comes from the poorly known $k_\perp$ dependence of the  sea-quark DPDs. In order to circumvent the problem, that, we emphasize, is not present for the factorized model of the nucleon DPDs,}
let us first define the following auxiliary ratio:

\begin{align}
   {\cal  R}_{A2}\left(x_1, x_2, {\bf k}_{\perp}\right)= \dfrac{D^{A,1}_{u_v u_v}\left(x_1, x_2, {\bf k}_{\perp}\right)}{D^{^2 H,1}_{u_v u_v}\left( x_1, x_2, {\bf k}_{\perp}\right)} ~ \dfrac{D^{^2 H,1}_{u_v u_v}\left(x_1, x_2, {\bf 0}_{\perp}\right)}{D^{A,1}_{u_v u_v}\left( x_1, x_2, {\bf 0}_{\perp}\right)}.
\end{align}

The above quantity can be formally used  {to recast the expression of}  the  {nuclear DPD} {to insert} in Eq. {\eqref{ratio1k}}, namely:

\begin{align}
    D^{A,1}_{u_v u_v}\left(x_1, x_2, {\bf k}_{\perp}\right) =  {\cal R}_{A2}\left(x_1, x_2, {\bf k}_{\perp}\right) D^{^2 H,1}_{u_v u_v}\left(x_1, x_2, {\bf k}_{\perp}\right) \dfrac{D^{ A,1}_{u_v u_v}\left(x_1, x_2, {\bf 0}_{\perp}\right)}{D^{^2 H,1}_{u_v u_v}\left( x_1, x_2, {\bf 0}_{\perp}\right)}
\end{align}
Therefore, the double-EMC ratio, Eq. \eqref{ratio1}, can be rewritten as follows:

\begin{align}
     R^{2,A,1}(z_1,z_2,x_1,x_2) &= \dfrac{D^{ A,1}_{u_v u_v}\left(x_1, x_2, {\bf 0}_{\perp}\right)}{D^{^2 H,1}_{u_v u_v}\left( x_1, x_2, {\bf 0}_{\perp}\right)} \dfrac{\displaystyle \int {d^2 {\bf k}_{\perp}\over (2 \pi)^2} 
 D^p_{\bar d \bar d}\left(z_1, z_2,- {\bf k}_{\perp} \right) D^{^2 H,1}_{u_v u_v}\left( x_1, x_2, {\bf k}_{\perp}\right) {\cal R}_{A2}\left(x_1, x_2, {\bf k}_{\perp}\right)   }{\displaystyle  \int {d^2 {\bf k}_{\perp}\over (2 \pi)^2} 
 D^p_{\bar d \bar d}\left(z_1, z_2,- {\bf k}_{\perp} \right) D^{^2 H,1}_{u_v u_v}\left( x_1, x_2, {\bf k}_{\perp}\right)   }
 \label{25}
 \\
 \nonumber
       &= \dfrac{D^{ A,1}_{u_v u_v}\left(x_1, x_2, {\bf 0}_{\perp}\right)}{D^{^2 H,1}_{u_v u_v}\left( x_1, x_2, {\bf 0}_{\perp}\right)}
      \Biggl[1-
       \int {d^2 {\bf k}_{\perp}\over (2 \pi)^2} 
 {\cal F}(z_1,z_2,x_1,x_2,{\bf k}_\perp)~ \Delta\left(x_1, x_2, {\bf k}_{\perp}\right)   \Biggr].
\end{align}
with

\begin{align}
\Delta\left(x_1, x_2, {\bf k}_{\perp}\right)= 1-{\cal R}_{A2}\left(x_1, x_2, {\bf k}_{\perp}\right)  
    \label{ratio3}
\end{align}
and

\begin{align}
    {\cal F}(z_1,z_2,x_1,x_2,{\bf k}_\perp)=\dfrac{D^p_{\bar d \bar d}\left(z_1, z_2,- {\bf k}_{\perp} \right) D^{^2 H,1}_{u_v u_v}\left( x_1, x_2, {\bf k}_{\perp}\right)}{S^{p ^2H,1}_{\bar d \bar d u_vu_v} (z_1,z_2,x_1,x_2)}.
\end{align}
 Recall that $ S^{p ^2H,1}_{\bar d \bar d u_vu_v} (z_1,z_2,x_1,x_2) >0$ and 
 
 \begin{align}
     \int {d^2 {\bf k}_{\perp}\over (2 \pi)^2}{\cal F}(z_1,z_2,x_1,x_2,{\bf k}_\perp) =1.
 \end{align}

  {We have  numerically checked for the RHC model that the quantity ${\cal R}_{A2}\left(x_1, x_2, {\bf k}_{\perp}\right)  $ for $^4$He differs from 1 less than 10\% in the region $x_1 + x_2 < 0.95$ when  ${\bf k}_\perp < 3$  GeV  for $x_2 = 0.3$ and $x_2 = 0.5$ (see Fig. \ref{fig:fatt}); analogous results can be obtained for other values of $x_2$.
  {We emphasize the region $x_1 + x_2 < 0.95$, since in the present paper we are mainly interested in the deep of the double-EMC ratio which is located in this range} {(see Figs. 2 and 3 of the main text)}. 
  Because of the Gaussian behavior of the DPDs in ${\bf k}_\perp$, the contribution to the integrals in Eq. \eqref{25} from values of ${\bf k}_\perp  $ larger than 3 GeV is negligible and therefore Eq. {{(12)}} of the main text holds at 10\% in the region $x_1 + x_2 < 0.95$, even for the RHC model where correlations are present.}
  {Finally, one can  conclude that the quantity:}

\begin{align}
     R^{2,A,1}(z_1,z_2,x_1,x_2) \to \frac{1}{Z} \dfrac{D^{ A,1}_{u_v u_v}\left(x_1, x_2, {\bf 0}_{\perp}\right)}{D^{^2 H,1}_{u_v u_v}\left( x_1, x_2, {\bf 0}_{\perp}\right)},
     \label{ratio2}
\end{align}
evaluated in the main text, is a good approximation of Eq. (\ref{ratio1})

\begin{figure}[t]
\includegraphics[width=0.45 \columnwidth]{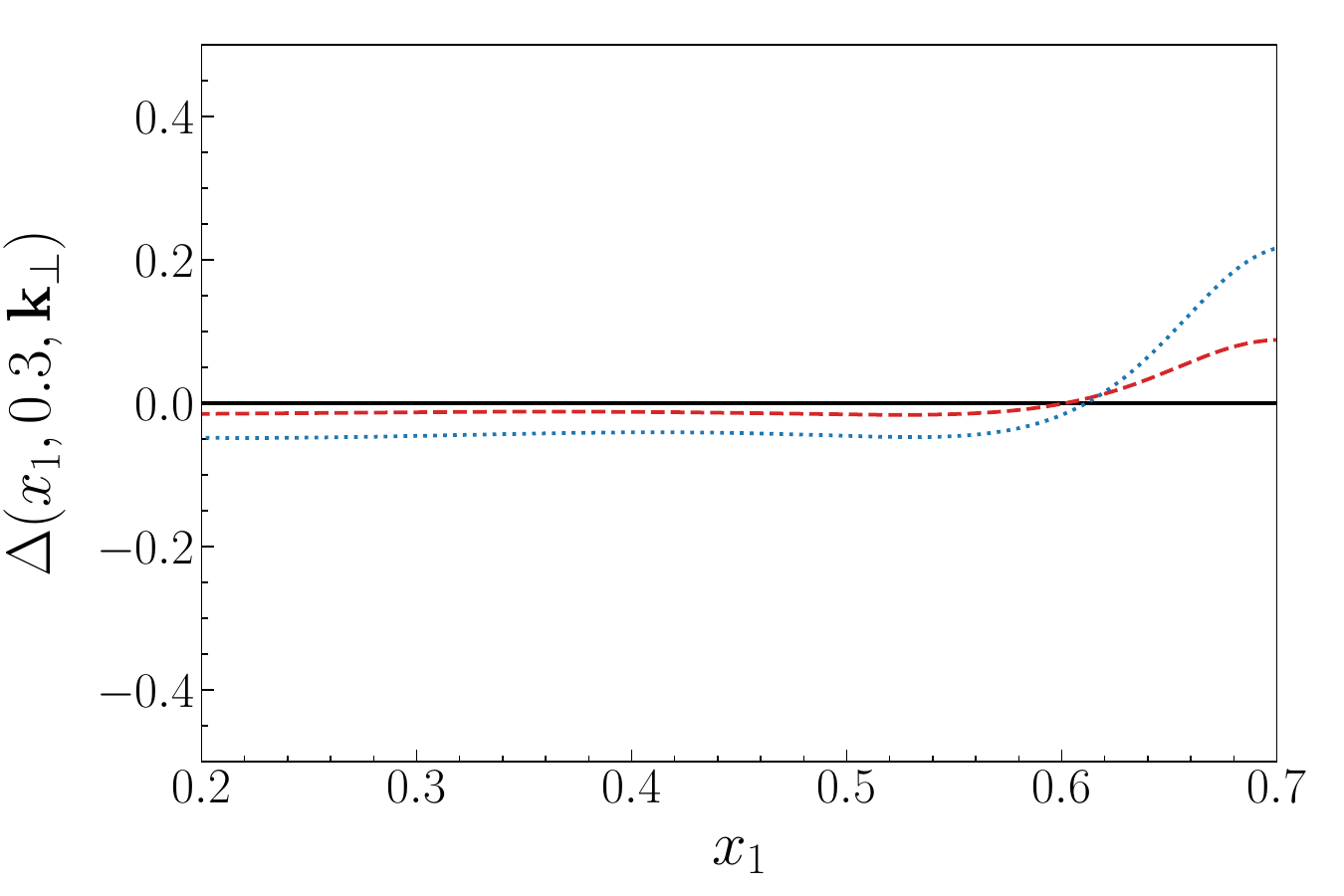}
\includegraphics[width=0.45 \columnwidth]{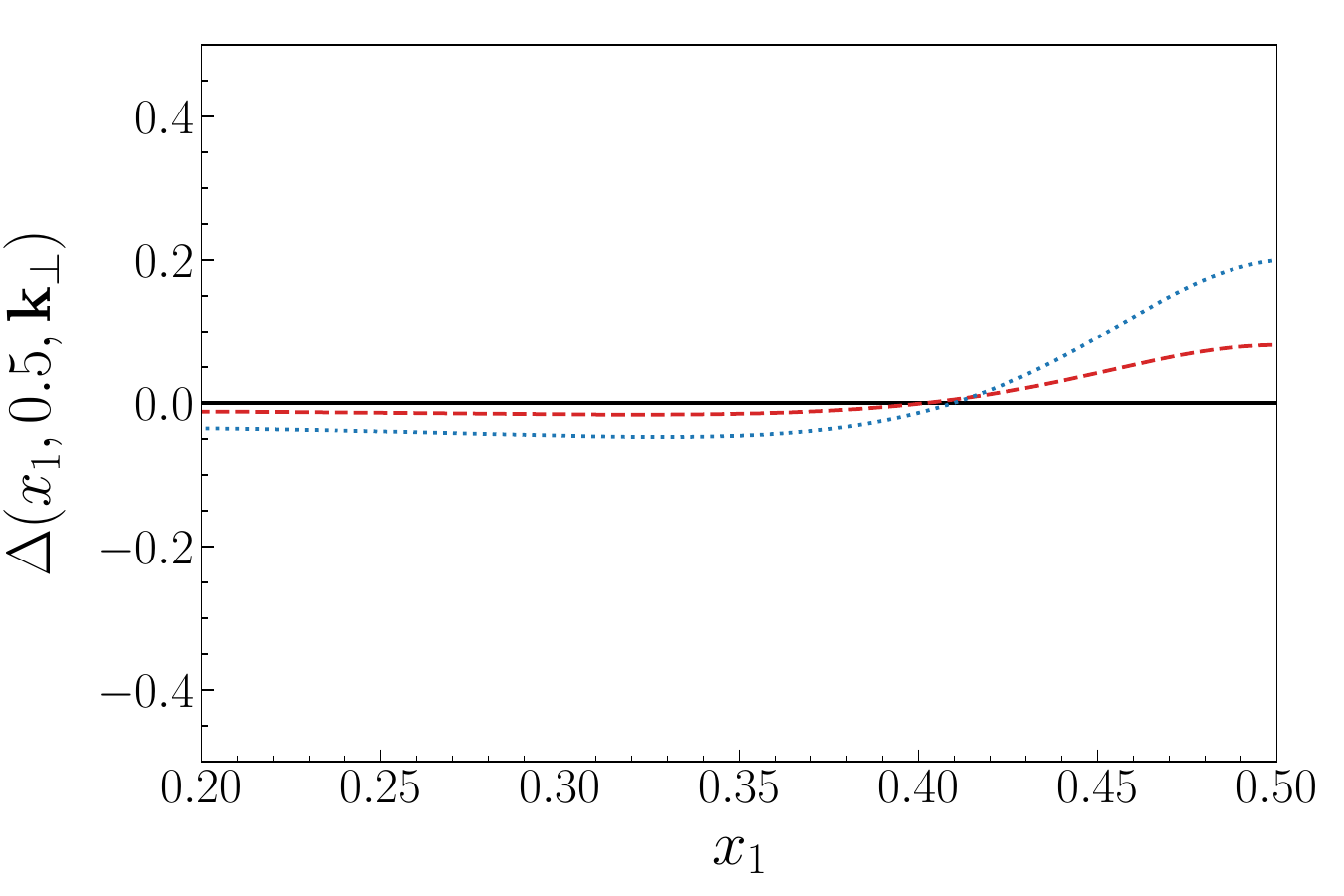}
\caption{ (Color online). Left panel:
The  difference $\Delta$ in Eq. \eqref{ratio3} evaluated at  $x_2=0.3$  for $k_\perp= 0$ GeV (solid line), $k_\perp= 1$ GeV (dashed line) and $k_\perp = 3$ GeV (dotted line). Right panel: the same of left panel but for $x_2=0.5$. 
}
\label{fig:fatt}
\end{figure}

\section{Bound nucleons} 
In this Section, we begin by  presenting  the approach adopted to modify bound nucleon PDFs and then two possible generalizations  for DPDs in nuclei, {when the factorized approximation of nucleon DPDs is assumed}. {It should be pointed out that we have chosen  not to make any modification to the correlated RHC model. }

 In our approach the nuclear LCMD, {which automatically fulfills both baryon and momentum sum-rules,} is defined as \cite{DelDotto:2016vkh}:

\begin{equation}
    \bar \rho^{A}_\tau(\xi) = A\int d \bold{k}_\perp n^A_\tau(\xi,\bold{k}_\perp),
\end{equation}
where $n^A_\tau(\xi,k_\perp)$ is the spin-independent nucleon momentum distribution obtained from the integration of the squared nuclear wave function $\mathcal{N}^A_\tau (\bold{k},\bold{k}_2, \dots, \bold{k}_{A-1})$ over the relative momenta of the remnant nucleons:
\begin{equation}    n^A_\tau(\xi,\bold{k}_\perp) = \frac{1}{2 \pi} \int \prod_{i=2}^{A-1} \big[ d \bold{k}_i\big] \left|\frac{\partial k_z}{\partial \xi} \right| \mathcal{N}^A_\tau (\bold{k},\bold{k}_2, \dots, \bold{k}_{A-1}),
\end{equation}
{Further details on the present approach can be found in Ref. \cite{Fornetti:2023gvf}.}
{The structure of nucleons in nuclei can be modified by the nuclear environment, {so that} nucleons are off their mass shell.}
Without off-shell (OS) effects, the nuclear SF is (see Ref. \cite{Pace:2022qoj}):
\begin{equation}
    {F}_2^A(x) = {\sum_{\tau=n,p}}\int_{\xi_{min}}^1 d \xi  F_2^\tau \left(\frac{x}{\xi} \bar{\xi} \right) \bar \rho^A_\tau(\xi) ,
    \label{eq:SFs_convolution}
\end{equation}
where  {$\xi_{min} = x\bar  \xi$ with $\bar \xi= m / M_A$,}  $m$ the nucleon mass  and $M_A$ the mass of the nucleus.
In order to add  OS effects to phenomelogical PDFs and SFs,  the model of Ref. \cite{Kulagin:2004ie} has been adopted. {Within this approach,} the modification of the PDFs is parametrized as the product of the virtuality of the nucleon {of momentum $p_\tau$}, $v^A_\tau = (p^2_\tau-m^2)/m^2$ {{times}} a function $\delta d (x)$ ({in the following we omit the dependence of $v_\tau^A$} on $\xi$ and $\bf k_\perp$):
\begin{equation}
    \tilde{d}_j^\tau(x) = d_j^\tau(x)\left[1 + v_\tau^A \delta d (x) \right]
    \label{OS_PDFs}.
\end{equation}
Since $\delta d (x)$  {is assumed independent of} the flavour of the quark, the modification of the SF has the same structure:
\begin{equation}
    \tilde{F}_2^\tau(x) = F_2^\tau(x)\left[1 + v_\tau^A \delta d (x) \right].
    \label{OS_SFs}
\end{equation}
The function $\delta d(x)$ is defined as follows \cite{Kulagin:2004ie}:
\begin{equation}
    \delta d (x) = C (x-\bar{x})(x-x_0)(1+x_0-x),
    \label{deltaf}
\end{equation}
where the three parameters $C$, $\bar{x}$ and $x_0$ are fitted to the data of the EMC effect. The parameter $C$ is related to the magnitude of the OS modification, while $\bar{x}$ and $x_0$ are the zeroes of $\delta d$, i.e. the values of $x$ where the SFs are unchanged by the OS effects. In our calculations we fitted only two parameters, $C$ and $x_0$, since the third one, $\bar x$, can be fixed by requiring that the modified PDFs still fulfill the baryon sum rule, i.e.:
\begin{equation}
    \int d x \, \big[\tilde{d}^p_{u_v}(x) +  \tilde{d}^p_{d_v}(x) \big] = \int d x \, \big[{d}^p_{u_v}(x) +  {d}^p_{d_v}(x) \big] = 3.
\end{equation}
From this requirement, one can easily obtain $\bar x$ as a function of the parameter $x_0$ as:
\begin{equation}
    \bar x = \dfrac{\int dx \, \Big\{\big[x^3 -x^2(2x_0 + 1) + x(x_0^2 + x_0) \big] \big[ {d}^p_{u_v}(x) +  {d}^p_{d_v}(x)\big]\Big\}}{\int dx \, \Big\{ \big[ x^2 - x(2 x_0 + 1) + x_0^2 + x_0\big] \big[ {d}^p_{u_v}(x) +  {d}^p_{d_v}(x)\big] \Big\}}
\end{equation}
This model has been adopted {to reproduce} the EMC ratio for $^4$He and $^3$He shown in Fig. 1 of the main text {in the case where the nucleon structure function of Refs. \cite{AUBERT1987740,SCHAFER1988175} is used}. {In order to take into account OS effects} in our Poincar\'e covariant approach for the nuclear structure function, we define a new quantity $V^{A}_\tau(\xi)$, representing the distribution of the virtuality of a nucleon $\tau$ with LC momentum $\xi$. This quantity {reads}
\begin{equation}
    V^{A}_\tau(\xi) = \int d \bold{k}_\perp v^A_\tau(\xi,\bold{k}_\perp) n^A_\tau(\xi,\bold{k}_\perp).
\end{equation}
{Hence}, by generalizing Eq. \eqref{eq:SFs_convolution}, the nuclear SF  is:
\begin{equation}
    \tilde{F}_2^A(x) = \sum_{\tau=n,p}\int_{\xi_{min}}^1 d \xi ~F_2^\tau \left(\frac{x}{\xi}\bar{\xi}\right)~\left[   \bar \rho^A_\tau(\xi) +  \delta d \left(\frac{x}{\xi} \bar{\xi} \right) V^{A}_\tau(\xi)\right]
\end{equation}
{and the {{corresponding}} modified nuclear PDF}:

\begin{equation}
    \tilde{d}_i^{A,OS}(x) = \sum_{\tau=n,p}\int_{\xi_{min}}^1 d \xi ~ \frac{\bar \xi}{\xi} d_i^\tau \left(\frac{x}{\xi}\bar{\xi}\right)~\left[   \bar \rho^A_\tau(\xi) +  \delta d \left(\frac{x}{\xi} \bar{\xi} \right) V^{A}_\tau(\xi)\right]
    \label{Eq:PDFOS}
\end{equation}

 {Since the parameters of the OS modification are fitted to the EMC data, our results are similar to those of Ref. \cite{Kulagin:2004ie}. {It is remarkable that,} in our framework, the LCMDs $\bar \rho^A_\tau(\xi)$ and the distribution of the virtuality $V^{A}_\tau(\xi)$  embed Poincar\'e covariant effects, which are not negligible {in the calculations of nuclear parton distributions}.}  
 In future study, we will address  the challenging problem of the structure of bound nucleons in a more systematic way,  specifically extending the calculations of the EMC effect to heavier nuclei (see {Ref. \cite{fornetti}}). 

We first {discuss the} fitting procedure for  the deuteron by considering the  ratio:
\begin{equation}
    R_2^{^2\text{H}}(x) = \frac{\tilde{F}_2^{^2H}(x)}{F_2^p(x) + F_2^n(x)}.
\end{equation}
{In Fig. \ref{fig:OS_Deu}, the  solid and dashed lines are the results {for $R_{2}^{^2\text{H}}(x)$} without and   with OS {effects}, respectively. The latter {results, in comparison with the BonUS data \cite{BonUS}  have been used to fix the OS parameters for the deuteron}. 
The deuteron $R_{2}^{^2\text{H}}(x)$ ratio is {then} used for computing the {EMC ratios}  for $^3$He and $^4$He:
\begin{align}
    & R_{EMC}^A(x) = \frac{R_2^A(x)}{R_{2}^{^2\text{H}}(x)} \end{align}
    with
    \begin{align}
     R_2^A(x) = \frac{\tilde{F}_2^A(x)}{Z F_2^p(x) + (A-Z)F_2^n(x)}.
\end{align}
In order to determine the only two OS parameters for $^3$He and $^4$He {EMC ratios, we  used}  the data of the E03-103 experiment \cite{Seely:2009gt},  as reanalized  in Ref. \cite{E03103_Re} for $^3$He. {The LCMDs obtained from the wave functions of Refs. \cite{Kievsky:1994mxj,Kievsky:1995uk} are used to evaluate the nuclear structure functions.} The results {for the EMC ratio}
{are} shown in Fig. \ref{fig:OS_A}. 
A full analysis including the details on the values of the parameters will provided in Ref. \cite{fornetti}.

\begin{figure} 
    \centering
    \includegraphics[width=0.45\columnwidth]{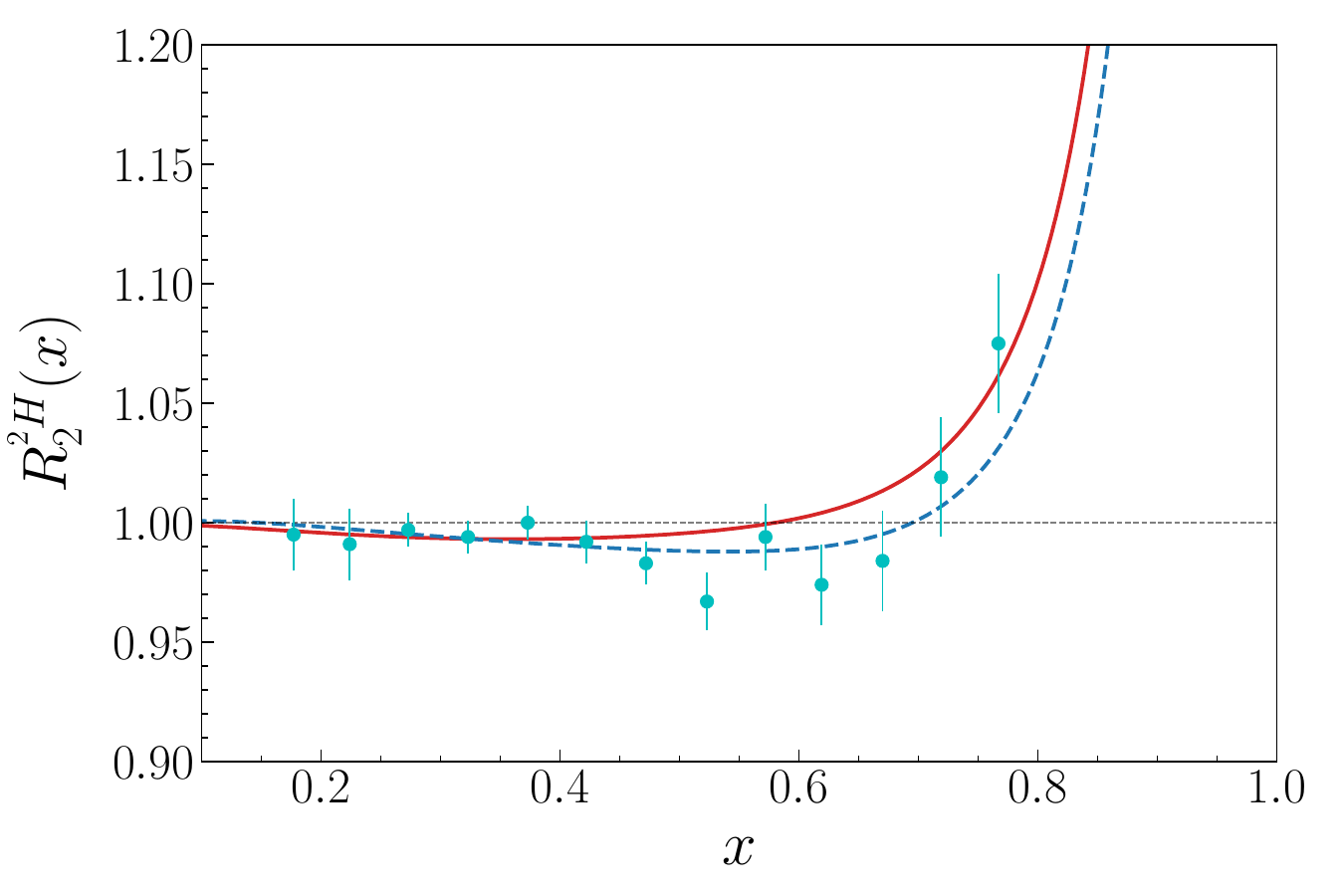}
    \caption{(Color online) The {${R_{2}^{^2\text{H}}(x)}$} ratio for the deuteron {computed} by using the EMPDF nucleon SFs \cite{AUBERT1987740,SCHAFER1988175}. 
    Solid line: results from Eq. \eqref{eq:SFs_convolution} without off-shell modification. Dashed line: calculation with OS {effects} from Eq. \eqref{deltaf} with  the parameters obtained by the fit to the data \cite{BonUS} extracted from the BonUS experiment \cite{BonUS1,BonUS2} }
    \label{fig:OS_Deu}
\end{figure}

\begin{figure}
    \centering
    \includegraphics[width=0.45\columnwidth]{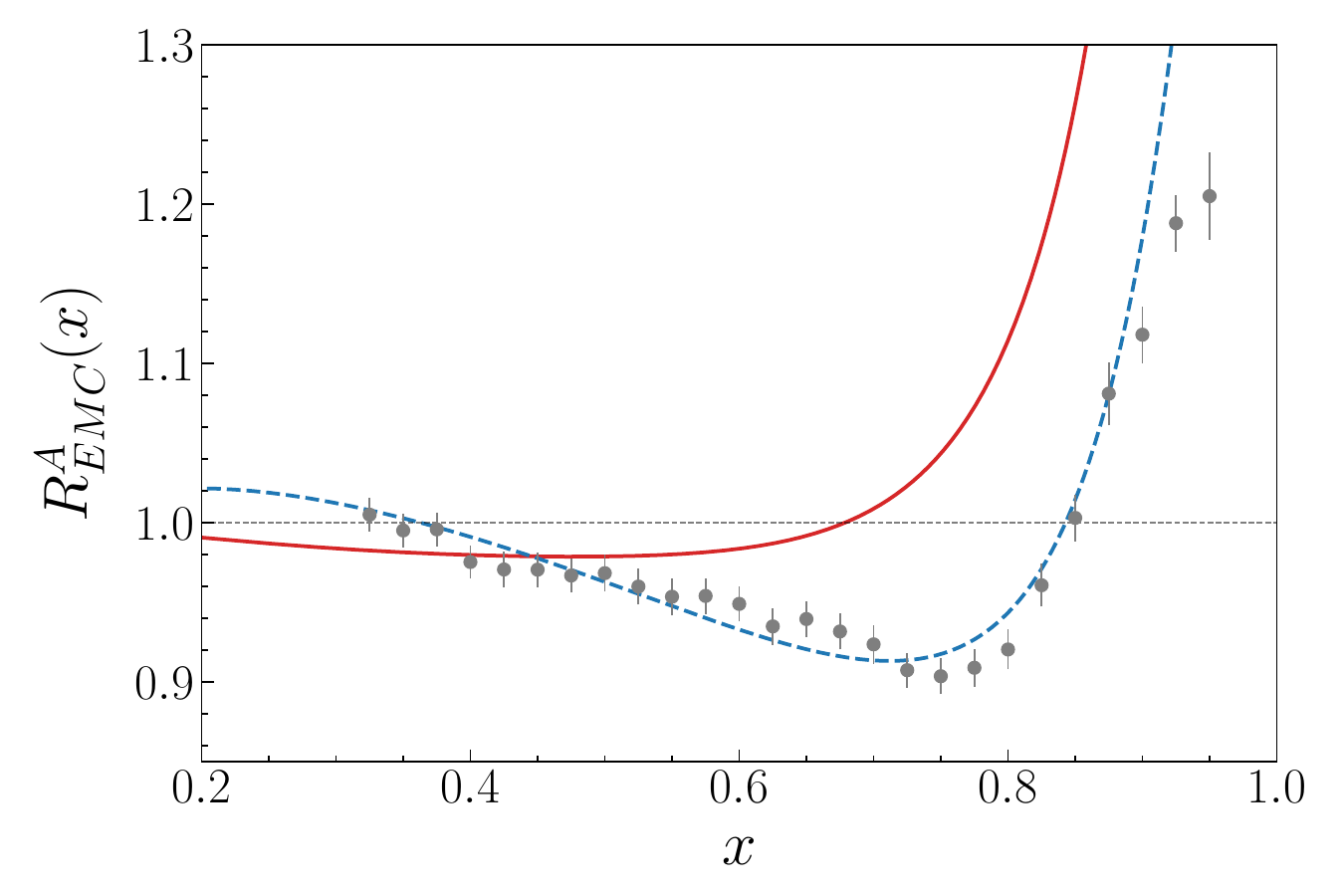}
    \includegraphics[width=0.45\columnwidth]{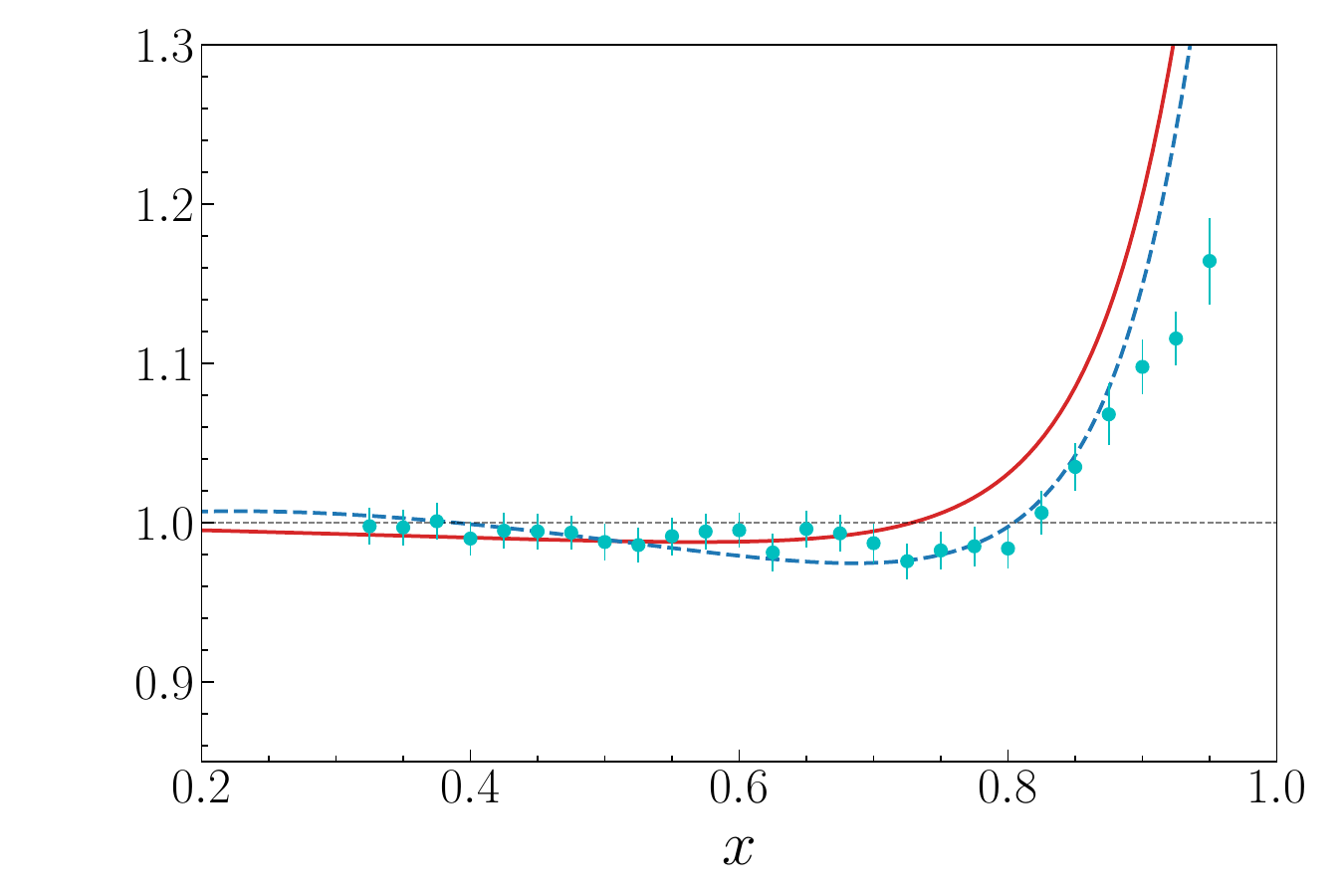}
    \caption{(Color online)  Left panel:  The EMC ratio for $^4$He computed by using the EMPDF nucleon SFs \cite{AUBERT1987740,SCHAFER1988175}. Solid line: EMC ratio {results} obtained by Eq. \eqref{eq:SFs_convolution} without off-shell modification. Dashed line: EMC ratio {with off-shell effects as given in Eq. \eqref{deltaf} with the parameters of Ref. \cite{fornetti}. Data from Ref. \cite{Seely:2009gt}}. Right panel: the same as the left panel, but for $^3$He.
 {The LCMDs obtained from the wave functions of Refs. \cite{Kievsky:1994mxj,Kievsky:1995uk}.} Data from Ref. \cite{Seely:2009gt}, as reanalized in Ref. \cite{E03103_Re}. }
    \label{fig:OS_A}
\end{figure}
 Our OS model has been {generalized} to evaluate  the nuclear DPDs in two different ways. In the first one, {that we call $OS1$,} {the factorized approximation of} the {nucleon} DPDs is modified by {adopting  an expression analogous to}  the nucleon one, Eq. \eqref{deltaf}, {but changing the argument {of $\delta d$} from $x$ to $x_1+x_2$, i.e.:} 
\begin{equation}
    \delta d_2(x_1 + x_2) = C_2 (x_1 + x_2-\bar{x}_2)(x_1 + x_2 - x_{0,2}) (1 + x_{0,2} - x_1 - x_2),
\label{DPD_OS1}\end{equation}
where $C_2 = C$ { has been implemented in order to keep the {same} magnitude of the OS effect found for the EMC ratios}.
Therefore, the {bound} nucleon DPD is given by:
\begin{equation}
    \tilde{{D}}^{\tau}_{ij}(x_1,x_2,\mathbf{0}_\perp) = {{D}}^{\tau}_{ij}(x_1,x_2,\mathbf{0}_\perp) \left[1 + v_\tau^A \delta d_2 (x_1+x_2) \right] ,
    \label{DPD_OS1b}
\end{equation}
and the nuclear  {DPD} is written as:
\begin{equation}
    {\tilde{D}}^{A,OS1}_{i j}(x_1,x_2,\mathbf{0}_\perp) = \sum_{\tau=n,p}\int_{\xi_{min}}^1 d \xi \, \frac{\bar \xi^2}{\xi^2} \left[ {{D}}^{\tau}_{ij}\left(\frac{x_1}{\xi} \bar{\xi},\frac{x_2}{\xi} \bar{\xi},\mathbf{0}_\perp\right) \bar \rho^A_\tau(\xi) + {D}^{\tau}_{ij} \left(\frac{x_1}{\xi} \bar{\xi},\frac{x_2}{\xi} \bar{\xi},\mathbf{0}_\perp\right)  \delta d_2 \left(\frac{x_1 + x_2}{\xi} \bar{\xi} \right) V^{A}_\tau(\xi)\right],
    \label{Eq:OS1}
\end{equation}
{where $\xi_{min} = (x_1 + x_2 )\bar \xi.$} }
The OS parameters can be extracted by  requiring that the modified DPDs still fulfills, {at the largest extent the following relation}:

\begin{align}
 \int_0^{1- x_1 \bar \xi/\xi}dy_2 ~
  {D}^{\tau}_{i j_v} \left(\frac{x_1}{\xi} \bar{\xi},y_2, {\bf 0}_\perp\right)  \delta d_2 \left(\frac{x_1 }{\xi} \bar{\xi}+y_2 \right) =   \delta d \left(\frac{x_1}{\xi}\bar \xi \right)
\begin{cases}\left(N^\tau_{j_{{v}}}-1\right) d_{i}^{A,\tau}\left(\frac{x_1}{\xi}\bar \xi \right) & i=j \\
N_{j_{{v}}}^{\tau} d_{i}^{A,\tau}\left(\frac{x_1}{\xi}\bar \xi \right) & i \neq j
\\
\left(N^\tau_{j_{{v}}}+1\right) d_{i}^{A,\tau}\left(\frac{x_1}{\xi}\bar \xi \right) & i=\bar j
\end{cases}
\label{Eq:OSSR}
\end{align}

{The above condition is sufficient to guarantee that the off-shell contribution in Eq. \eqref{Eq:OS1} does not spoil the nuclear GS sum rule Eq. \eqref{NSR_DPS0}, for $\tilde D^{A,OS}_{ij}(x_1,x_2, {\bf 0}_\perp)$. It is worth mentioning that if the free nucleon DPDs do not completely  fulfill the GS sum rules, as in the factorized Ansatz, then the corresponding  nuclear DPDs will numerically break these constraints.   }

In the second approach, {called OS2}, { the OS effect is implemented at the level of the single PDFs, so that}, we {insert Eq. \eqref{OS_PDFs} in Eq. \eqref{DPDs_cambio_variabile}} {obtaining}:
\begin{equation}
    \tilde{{D}}_{ij}^{\tau,OS2}(x_1,x_2,\mathbf{0}_\perp) = {{D}}_{ij}^{\tau}(x_1,x_2,\mathbf{0}_\perp) + v_\tau^A \delta G_1^\tau(x_1,x_2) + \big(v_n^A\big)^2 \delta G_2^\tau(x_1,x_2),
    \label{OS_SFs}
\end{equation}
with:
\begin{equation}
\begin{split}
    \delta G_1^\tau(x_1,x_2) =  &\frac{\lambda_{i j}}{2}  \left[ \frac{1}{1-x_2} d_i^\tau\left(\frac{x_1}{1-x_2}\right)d_j^\tau(x_2) \left[ \delta d\left( \frac{x_1}{1-x_2}\right) + \delta d(x_2)\right]+ \right.
    \\ 
    & \left. +\frac{1}{1-x_1} d_j^\tau(x_1) d_i^\tau \left(\frac{x_2}{1-x_1}\right) \left[ \delta d\left( \frac{x_2}{1-x_1}\right) + \delta d(x_1)\right]  \right] \theta(1-x_1 - x_2)
\end{split}
\end{equation}
\begin{equation}
\begin{split}
    \delta G_2^\tau(x_1,x_2) =  &\frac{\lambda_{i j}}{2}  \left[ \frac{1}{1-x_2} d_i^\tau\big(\frac{x_1}{1-x_2}\big)d_j^\tau(x_2) \big[ \delta d\big( \frac{x_1}{1-x_2}\big) \delta d(x_2)\big] \right. +\\ 
    & \left. +\frac{1}{1-x_1} d_j^\tau(x_1) d_i^\tau \left(\frac{x_2}{1-x_1}\right) \left[ \delta d\left( \frac{x_2}{1-x_1}\right) \delta d(x_1)\right]  \right] \theta(1-x_1 - x_2).
\end{split}
\end{equation}
{We recall that $d\delta(x_i)$ is given in Eq. \eqref{deltaf} and the parameters therein  are fixed to reproduce the EMC effect for the given nucleus. }
{In} this approach {is present} a term proportional to $(v_A^\tau)^2$, {thus} a new distribution {is introduced, i.e.}:
\begin{equation}
  V^{A,2}_\tau(\xi) = \int d \bold{k}_\perp \big[v^A_\tau(\xi,\bold{k}_\perp)\big]^2 n^{A}_\tau(\xi,\bold{k}_\perp),
\end{equation}
{and therefore the DPDs can be rewritten} as follows:
\begin{equation}
    \tilde{{D}}_{ij}^{A,OS2}(x_1,x_2,\mathbf{0}_\perp) = \sum_{\tau=n,p}\int_{\xi_{min}}^1 d \xi \left[ \tilde{{D}}_{ij}^{\tau}\left(\frac{x_1}{\xi} \bar{\xi},\frac{x_2}{\xi} \bar{\xi},\mathbf{0}_\perp\right) \bar\rho^A_\tau(\xi) + \delta G_1^\tau \left(\frac{x_1}{\xi} \bar{\xi},\frac{x_2}{\xi} \bar{\xi} \right)V^{A}_\tau(\xi) 
     + \delta G_2^\tau \left(\frac{x_1}{\xi} \bar{\xi},\frac{x_2}{\xi} \bar{\xi}\right) V^{A,2}_\tau(\xi)\right].
    \label{Eq:OS2}
\end{equation}

Finally, we  {{checked}}  the GS number sum rules for the DPDs  {{with the factorization Ansatz and OS effects}} for both  the {models} and compared to the results of GS number sum rules without OS effects. {To this aim we {{studied for  $^4$He the validity of}} the following relation, from Eq. \eqref{NSR_DPS0}:}

\begin{align}
    \frac{x_1}{2} \int^{A-x_1}_0 dx_2 \Big[ D^A_{u_v u_v}(x_1,x_2,{\bf 0}_\perp)+ D^A_{u_v d_v}(x_1,x_2,{\bf 0}_\perp) \Big] = x_1 d^A_{u_v}(x_1) 
     \label{Eq:test}
\end{align}
{{while in}} the case of presence of OS effects we have {to study the validity of the following equation}:

\begin{align}
    \frac{x_1}{2} \int^{A-x_1}_0 dx_2 \Big[ \tilde D^{A,OS1(2)}_{u_v u_v}(x_1,x_2,{\bf 0}_\perp)+ \tilde D^{A,OS1(2)}_{u_v d_v}(x_1,x_2,{\bf 0}_\perp) \Big]  = x_1 \tilde  d^{A,OS}_{u_v}(x_1). 
    \label{Eq:testOS}
\end{align}
In the left panel of Fig. \ref{fig:OS_SR} we report the {{results for}} Eq. \eqref{Eq:test}, {where OS effects are not included.  The factorization Ansatz of the free nucleon DPD leads to a violation of the sum rule Eq. \eqref{NSR_DPS0}.
In the right panel of Fig. \ref{fig:OS_SR}, we compare the 
left {hand} side of Eq. \eqref{Eq:testOS} obtained within the OS1 e OS2 models with the {{results for}} the
right {hand} side of the same equation. One should notice that $i)$ the violation of  Eq. \eqref{Eq:testOS} for the OS2 approach is of the same order of that shown in the left panel of Fig. \ref{fig:OS_SR} in absence of OS effects; $ii$) in the case of  OS1, the parameters encoded in $\delta d_2$ and fixed to fullfill Eq. \eqref{Eq:OSSR}, lead to same violation observed {{without the OS effects}}, as expected. In closing, we found that the overall error on the baryon sum rule,   induced by the factorization Ansatz, is around $7 \%$.   }

\begin{figure}
    \centering
    \includegraphics[width=0.45\columnwidth]{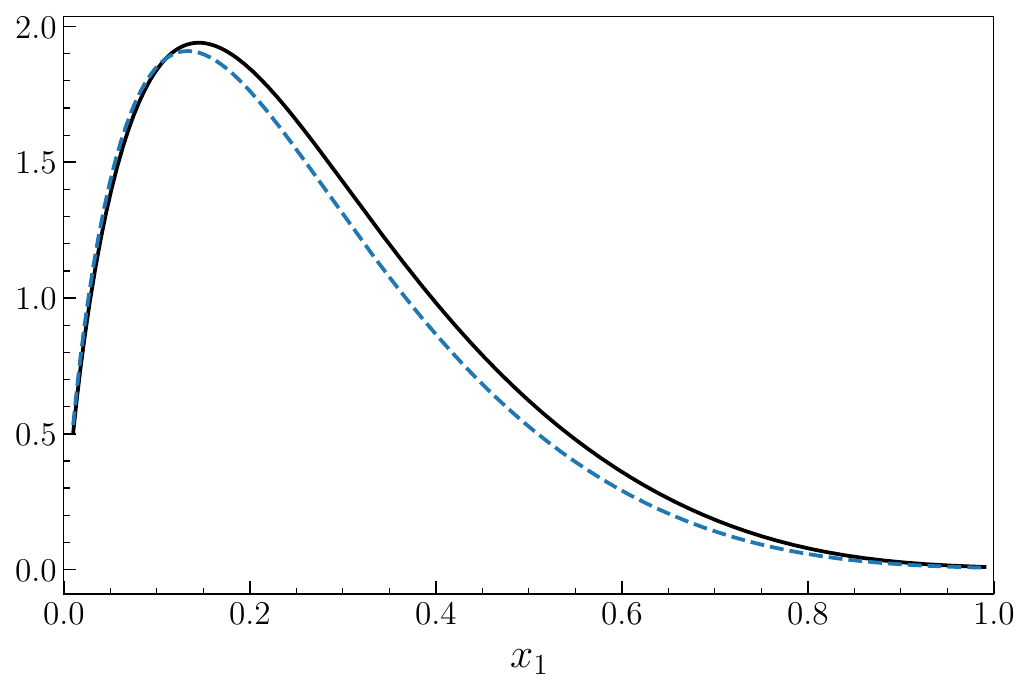}
\includegraphics[width=0.45\columnwidth]{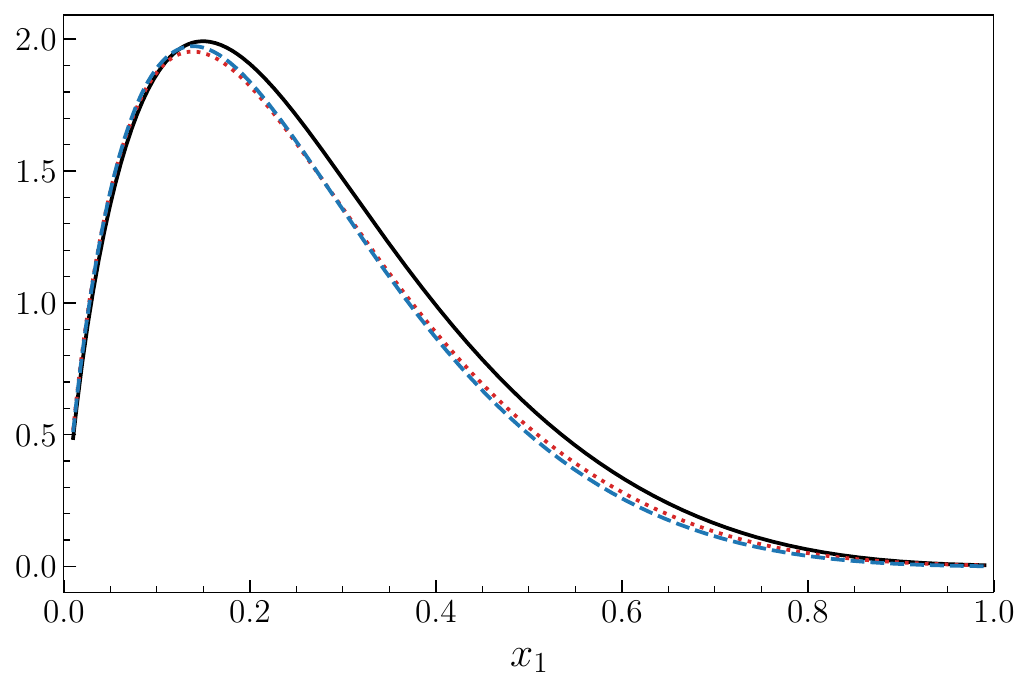}
    \caption{(Color online)  Left panel: 
    the quantity $x_1 d^{^4 He}_{u_v}(x_1)$ (full line), right {hand} side of Eq. \eqref{Eq:test}, compared with the evaluation of the left {hand} side of the same equation  (dashed line).
    Right panel: the quantity $x_1 \tilde d^{^4 He,OS}_{u_v}(x_1)$ (full line), right {hand} side of Eq. \eqref{Eq:testOS}, compared with the evaluation of the left {hand} side of the same equation (dotted line for OS1 and dashed line for OS2).}
    \label{fig:OS_SR}
\end{figure}

\bibliographystyle{elsarticle-num} 
\bibliography{bib3}






\end{document}